# Pawnshops and Property-Related Crime: Lessons from Interwar Tokyo

Tatsuki Inoue*

Abstract

Using ward-level panel data for 1927 to 1935 and fixed-effects models, this study examines how pawnshops were related to property-related crime in interwar Tokyo. Accordingly, public pawnshops targeting low-income borrowers are compared with profit-oriented private pawnshops. The results reveal that public lending is negatively associated with theft and positively associated with fraud and extortion, whereas private lending is negatively associated with fraud and extortion as well as embezzlement. A simple monetization exercise shows that offense counts and victim losses can yield different assessments. Overall, the relationship between credit and crime depended on both institutional design and crime type.

* School of Commerce, Meiji University, Faculty Office Building, 1-1, Kanda-Surugadai, Chiyoda-ku, Tokyo, 101-8301, Japan. E-mail: tatsukiinoue@meiji.ac.jp.

# 1. Introduction

Does fringe banking increase or reduce crimes involving the unlawful acquisition of money or other property? Fringe banking, including pawnshops, provides small, short-term credit outside the banking sector and may be linked to such crimes in opposing ways. On the one hand, short-term credit may reduce incentives to commit theft by easing sudden liquidity shortages and temporary income shocks (McCannon et al. 2022; Morse 2011). Conversely, relatively high-cost credit may create new repayment burdens and thereby increase illegal attempts to obtain funds (Aaltonen et al. 2016; Melzer 2011). Moreover, this type of credit may be connected to crime not only through borrowers' own behavior but also through the opportunity structure surrounding borrowers and lending outlets (d'Este 2020; Kubrin and Hipp 2016). Theoretically, the relationship between fringe banking and property-related crime is therefore not determined in a single direction.

Likewise, empirical research has offered no consistent conclusions on the relationship between fringe banking and theft-related crime. Some studies argue that access to such credit increases crime (Kubrin et al. 2011), others associate it with reductions in crime (Barth et al. 2020; McCannon et al. 2022), and still others find no clear relationship (Lee et al. 2014). These mixed findings may reflect insufficient attention to differences in institutional design and primary user groups. In addition, the relationships with property-related crimes other than theft, such as fraud and embezzlement, remain understudied.

To address this knowledge gap, this study focuses on pawnshop credit as a form of fringe banking and uses ward-level panel data for the 15 wards of Tokyo from 1927 to 1935. Following the conventional Japanese classification of offenses against property, this study treats theft, fraud/extortion, and embezzlement as property-related offenses and examines how the relationship between pawnshop credit and property-related crime varies by crime type. Interwar Tokyo also had both public pawnshops, which primarily targeted low-income borrowers, and private pawnshops, which operated for profit and served a broader clientele. The case of interwar Tokyo therefore allows this study to examine both heterogeneity within property-related crime and institutional differences in pawnshop credit. This period and setting are also well suited to examining the relationship between pawnshop credit and crime rates, as they were marked by substantial social uncertainty generated by economic shocks.

Empirically, this study estimates the association between public and private pawnshop lending and crime rates using fixed-effects models, with primary focus on heterogeneity of associations across crime types and the contrast between public and private pawnshops. Because time-varying unobserved factors and reverse causality cannot be fully ruled out, the estimated coefficients are interpreted as conditional associations rather than strict causal effects.

The results indicate that the relationship between pawnshop credit and property-related crime is not uniform. Increases in public pawnshop lending are negatively and positively associated with theft rates and fraud/extortion rates, respectively, but display no clear relationship with embezzlement rates. By contrast, private pawnshop lending exhibits no clear relationship with theft rates but is negatively associated with fraud/extortion and embezzlement rates. These differences are difficult to detect when property-related crime is considered as a single aggregate category and become clear only when such crime is divided into crime types.

The variation in results across combinations of pawnshop and crime types suggests that the relationship between pawnshop credit and property-related crime cannot be explained by a single mechanism. The relative balance of these mechanisms depends on both institutional design and crime type. The conceptual framework presented below develops these mechanisms to interpret why public and private pawnshops may be associated with different types of crime. Although this study does not rigorously identify each potential channel, these channels help interpret why public and private pawnshops were associated with different types of crime.

A simple back-of-the-envelope monetization exercise further demonstrates that these heterogeneous associations have different implications when evaluated in terms of recorded offense counts and victim losses. In particular, public pawnshops appear to have been associated with fewer crimes but higher, rather than lower, victim losses from property-related crime. By contrast, the private-pawnshop results imply sizable reductions in victim losses because the negative associations are concentrated in higher-loss crime types.

This study makes several contributions to the literature. First, the relationship between pawnshop credit and property-related crime differed across crime types. Previous studies of fringe banking and crime have largely focused on theft-related outcomes, such as burglary, larceny, motor-vehicle theft, and related theft-type offenses (Barth et al. 2020; Kubrin et al. 2011; Lee et al. 2014; McCannon et al. 2022). This study adds findings from historical Japan to that literature and extends the analysis beyond theft-related outcomes to fraud/extortion and embezzlement. These differences indicate that evaluations based only on theft-related outcomes or an aggregate measure of these property-related crimes can miss important heterogeneity.

Second, the study reveals that even within pawnshop credit, public and private pawnshops were associated with different types of property-related crime. This finding relates to research on fringe banking indicating that the institutional design of small, high-cost credit, including interest-rate ceilings, loan maturities, screening, and supply restrictions, can shape borrowing behavior and social outcomes (Carter et al. 2022; Caskey 1991; Melzer 2011; Shackman and Tenney 2006; Stegman 2007). The present study also connects to research on the institutional design of microcredit, which demonstrates that repayment schedules and liability structures can affect borrower outcomes (Field and Pande 2008; Field et al. 2013; Giné and Karlan 2014). This study extends the importance of institutional design to property-related crime as an outcome, further revealing that even the same form of pawn credit may be associated with different social implications depending on institutional arrangements.

The findings on the association between pawnshops and crime rates also recast pawnshops in interwar Tokyo not merely as alternative financial institutions but as institutions that may have been connected to urban order or disorder. The historical literature on private pawnshops has mainly investigated the institutional and market structure of pawnbroking, the maintenance of everyday life among the urban lower classes, and risk coping in response to household shocks (e.g., Inoue 2021; Ioku and Shizume 2014; McLaughlin and Pecchenino 2022; Murhem 2016; Shibuya et al. 1982; Woloson 2009). In addition, recent historical studies have reassessed public pawnshops, or charitable pawnbroking, as institutions of poor relief, urban popular finance, and welfare provision (Carbonell-Esteller 2012; Carboni and Fornasari 2019; Inoue 2023; McLaughlin 2013; Pastureau and Blancheton 2014).[1] However, the broader social consequences beyond borrower welfare remain insufficiently examined in the literature. By linking pawnshop credit to property-related crime and victim losses, the present study offers a perspective for reconsidering the historical significance of pawnshops in relation to urban order.

The remainder of the paper is organized as follows. Section 2 describes property-related crime and pawnshop institutions in interwar Tokyo. Section 3 develops the conceptual framework linking pawnshop credit to different types of property-related crime. Section 4 explains the data and empirical strategy. Section 5 presents the main results and additional analyses. Section 6 discusses the substantive magnitude and monetary implications of the estimates. Section 7 concludes.

[1] The role of public pawnshops may have differed across regions and periods. Based on daily ledgers from a Swedish pawnshop from 1880 to 1950, Murhem and Ulväng (2023) argue that to regard the typical user as belonging to the very poorest strata is inappropriate.

## 2. Historical Background

### 2.1. Crime Trends and Property-Related Crime in Interwar Tokyo

From the late 1920s to the early 1930s, the Japanese economy entered a period of severe stagnation following the financial crisis and the Showa Depression. This period was characterized by recession, business failures, rising unemployment, and widespread hardship among the population (Research and Training Institute, Ministry of Justice, ed. 1960). Given the high concentration of commercial functions and population, as well as the large numbers of precariously employed and low-income residents, this deterioration in economic conditions likely had a strong effect on Tokyo. According to the 1930 Population Census, the number of unemployed people in Tokyo reached 22,271 (Cabinet Statistics Bureau 1933). A survey conducted between October 1932 and February 1933 also reported 48,890 protection-requiring households (Tokyo City Social Welfare Bureau 1933).[2]

[2] For living conditions among the lower classes in prewar Tokyo, see Nakagawa (1985).

Crime was also entering an expansionary phase nationwide during the same period. The number of Penal Code offenses reportedly began to increase overall around 1929 and peaked in 1933–1935 (Research and Training Institute, Ministry of Justice, ed. 1960). The number of cleared cases also rose from 1928 to 1940 and reached its prewar high in 1934 (Research and Training Institute, Ministry of Justice, ed. 1968).[3]

Tokyo exhibited a similar increase in crime rates during this period. Trends in major crime rates in Tokyo from 1920 to 1940 are illustrated in Figure 1. After the Great Kanto Earthquake of 1923, the total crime rate in Tokyo began to rise around 1927, increasing from roughly 28 cases per 1,000 populations to approximately 45 cases by 1933, before declining to approximately 38 cases by 1935. Therefore, the study period, from 1927 to 1935, precisely captures the phase in which Tokyo's crime rate rose, peaked, and then reversed.

In Figure 1, another important point was that crime rates in Tokyo during this period were composed primarily of property-related crimes. Theft accounted for slightly more than three quarters of all recorded crimes. Although fraud/extortion and embezzlement were smaller in scale, these three types combined constituted the dominant components of recorded crime (Tokyo Metropolitan Police Department 1934). These property-related crimes were therefore not peripheral to Tokyo's public-order environment, but were central to it.

---

[3] The sharp increase in cleared cases around 1934 may partly reflect intensified police enforcement and the discovery or processing of cases from previous years (Research and Training Institute, Ministry of Justice, ed. 1968). In addition, as institutions and police enforcement practices changed over time, strict time-series comparisons require caution. Regarding changes in the relationship between the police and the public in prewar Japan, see Obinata (1993).

At the same time, these three types of property-related crime did not move uniformly. Theft drove the overall trend, while embezzlement was smaller in scale and more stable. Long-run national statistics also indicate that these crime types followed distinct trajectories (Research and Training Institute, Ministry of Justice, ed. 1960), suggesting that the rise in crime rates contained internal differences that cannot be adequately captured by aggregate property-related crime alone.

Tokyo Metropolitan Police Department (1934) statistics that classify arrestees in 1933 by stated cause further indicate that the backgrounds of these property-related crimes were not necessarily identical. For theft, unemployment and poverty together accounted for approximately 14 percent of arrestees. Although greed, idleness, and impulse were more frequently cited at approximately 36, 17, and 22 percent, respectively, these categories are highly ambiguous and may have included other underlying motives. Hence, the fact that roughly 15 percent of arrestees explicitly cited economic circumstances highlights the connection between theft and livelihood hardship. For fraud, greed accounted for nearly half of arrestees, with unemployment and poverty combined at roughly 11 percent. For embezzlement, economic causes accounted for only a small share, less than 10 percent. The weight of motives differed across crime types, suggesting heterogeneity within property-related crime.

Victim-side statistics from the Statistical Reports of the Tokyo Metropolitan Police Department for 1933 further reinforce this heterogeneity. Among victims of fraud and extortion, approximately 34 percent were recorded as having no occupation or an unknown occupation, compared with approximately 10 and 16 percent among victims of theft and embezzlement, respectively. These figures indicate that fraud and extortion were more closely associated than theft and embezzlement with victims whose occupational status was absent or difficult to classify, suggesting that these offenses depended more heavily on victims' characteristics and vulnerability.

The geographical distribution of crime rates across the 15 wards from 1927 to 1935 is presented in Figure 2a. Crime rates varied substantially across wards; some in the central and eastern parts of the city had relatively high crime rates, while parts of the western and southern areas remained at lower levels. Because the city center was a high-income area whereas the eastern wards were low-income areas, the distribution indicates that crime rates did not depend mechanically on ward-level income.

## 2.2. Pawnshops in Interwar Tokyo

Pawnshops played an important role in modern Japan as providers of small, short-term secured credit.[4] In 1895, the Pawnshop Control Law (*Shichiya Torishimari Ho*) was enacted. Under this law, private pawnshops were subject to interest-rate ceilings and lending rules, and were placed under a licensing system intended to facilitate investigations of stolen goods and regulate business operations. In 1927, the Public Pawnshop Law (*Koeki Shichiya Ho*) was enacted, which established the legal foundation for public pawnshops. During the period analyzed in this study, Japan's pawnbroking system therefore consisted of two types of pawnshops: private pawnshops operating under the existing regulatory law and public pawnshops developing under a separate legal framework.

Tokyo was a city in which these pawnshops had an especially prominent place in urban finance.[5] In interwar Tokyo, private pawnshops continued to operate widely, whereas public pawnshops were also established by Tokyo. The latter increased from eight branches in 1927 to 17 in 1930, whereas private pawnshops still operated on a scale of roughly 700 establishments. This indicates that the development of public pawnshops did not eliminate private pawnshops; rather, two types of pawnshop credit coexisted within the same city. Urban residents in interwar Tokyo therefore faced not a single homogeneous pawnbroking market, but rather a credit environment in which two institutionally distinct types of pawnshops operated side by side.

---

[4] Tokyo City Office (1931) explained that among financial institutions for ordinary people, including savings banks, credit cooperatives, mutual loan associations, moneylenders, and pawnshops, only pawnshops could provide living funds quickly and easily, making them the most representative institution.

[5] For the historical development of pawnshops in interwar Tokyo, see Inoue (2023). Shibuya et al. (1982) provide a particularly detailed account of the enactment and revision of both laws.

However, public pawnshops were not distributed citywide; the geographical distribution of pawnshops is presented in Figure 2b. The shading of each ward represents the distribution of private pawnshops, while the numbers indicate the number of public pawnshops. The latter were located in a limited number of wards and were especially concentrated in eastern wards such as Shimoya, Asakusa, Honjo, and Fukagawa. By contrast, private pawnshops were distributed widely across Tokyo, although their density varied across wards. Notably, the concentration of public pawnshops in low-income eastern wards does not simply coincide with the geography of crime (Figure 2a).

Pawnshops were not peripheral institutions used only by a small group of exceptional borrowers. A 1933 survey by the Tokyo Prefecture Department of Academic Affairs (1935) indicates that 73 percent of borrowing by poor households was conducted through pawnshops. In addition, 99 percent of pawned items were clothing, indicating that collateral in pawnshop transactions did not consist of special high-value assets. The main use of borrowed funds was living expenses, which accounted for as much as 85 percent of the total. These facts demonstrate that pawnshops functioned as everyday small-scale financial institutions tied to the maintenance of livelihoods among low-income households. As emphasized by many contemporary sources, pawnshops were the institution most closely connected to the cash management of urban residents (Social Welfare Bureau 1926; Tokyo City Office 1931; Tokyo Institute for Municipal Research 1925).

### 2.3. Institutional Differences between Public and Private Pawnshops

Public and private pawnshops shared the basic feature of providing loans secured against pawned items. Under both systems, users could borrow small amounts of money by pledging clothing or other goods as collateral. However, the two types of pawnshops differed clearly in institutional purpose. Public pawnshops were designed to provide funds to economically distressed individuals on relatively favorable terms, whereas private pawnshops were profit-oriented private financial businesses. Contemporary observers also understood public pawnshops as institutions that linked borrower welfare to broader social stability.[6] The distinction between pawnshops was therefore not merely a difference in ownership; it concerned the question of to whom, and under what conditions, credit was supplied.

This difference was first clearly reflected in interest rates. Private pawnshops charged monthly interest rates of 2.5 to 4.0 percent, whereas the rate for public pawnshops was limited to 1.25 percent per month.[7] Institutionally, therefore, the interest rate charged by public pawnshops was set at less than half that charged by private pawnshops. Because the latter tended to apply relatively higher rates to smaller loans (Shibuya et al., 1982), this difference was especially important for low-income borrowers who needed small amounts of credit. Thus, although both public and private pawnshops operated through the same basic form of pawnbroking, they generated substantial differences in the cost of borrowing faced by users.

---

[6] A contemporary employee of a public pawnshop described the institution as a unique social-policy agency with responsibility for both promoting users' welfare and maintaining social peace (Koukei 1928).

Differences in interest burdens may have been partly reflected in pawn forfeiture. The forfeiture rate among completed transactions at the ward-year level from 1927 to 1935 is presented in Figure 3a. The rate for public pawnshops was distributed at a relatively low level overall, with a mean of 5.7 percent, while that for private pawnshops was clearly located at a higher level. The center of the distribution was roughly between 9 and 13 percent, with a mean of 11.9 percent. Thus, forfeiture rates differed clearly by institutional type.

The differences between public and private pawnshops extended beyond interest rates to loan ceilings and actual loan sizes. The Public Pawnshop Law imposed loan ceilings on public pawnshops of 10 and 50 yen per loan and per household, respectively.[8] This restriction made it possible to limit the inflow of middle-income users and borrowers with relatively large financing needs. Consequently, while private pawnshops served customers up to the middle-income class, including merchants and salaried employees, the role of public pawnshops was confined to lending for low-income groups (Tokyo City Social Welfare Bureau 1926). According to Tokyo City Office (1926), loans exceeding 10 yen accounted for approximately 25 percent of all loans at private pawnshops. By contrast, loans at public pawnshops were smaller, with about 64 percent of all loans amounting to 5 yen or less. The distinction between pawnshops was not only limited to differences in interest burdens but also concerned what types of borrowers each institution served and what scale of short-term funds it supplied.

[7] The annual bank lending rate in Tokyo was 2.12 percent in 1930 (Bank of Japan 1986); thus, even the rate charged by public pawnshops was high relative to bank credit, which was effectively limited to high-income borrowers.

Differences in loan size are also observable at the ward level. The average loan amount per transaction is presented in Figure 3b, with those at public pawnshops concentrated at relatively low levels. The main mass of the distribution lay between the 4- and 5-yen ranges, thus indicating that lending was generally skewed toward small loans. By contrast, average loan amounts at private pawnshops were larger overall than those at public pawnshops. The center of the distribution lay around 5 to 8 yen. The distribution also had a long right tail, with some observations exceeding 10 yen. This distribution indicates that public pawnshops provided smaller and relatively homogeneous loans, whereas private pawnshops provided larger loans and displayed wider variation in loan size.

[8] Tokyo changed the maximum loan amount to 20 and 100 yen per loan and per household, respectively, in 1930 (Tokyo Prefecture Department of Academic Affairs, 1935).

## 3. Conceptual Frameworks

### 3.1. Potential Channels Linking Pawnshop Credit to Property-Related Crime

This section organizes the potential channels through which pawnshop lending may affect property-related crime rates. Drawing on the economic theory of crime (Becker 1968; Ehrlich 1973), the analysis is premised on the fact that access to credit can alter both participation in illegal activities and the opportunity structure for crime. Short-term credit, such as pawnshop lending, may reduce crime arising from income shocks but may also increase property-related crime through other channels, thereby making the net theoretical prediction ambiguous.

The first channel is the relaxation of liquidity constraints. Household shocks, such as unemployment and income loss, are widely recognized to increase participation in crime (Blakeslee et al. 2018; Britto et al. 2022; Khanna et al. 2021), and access to credit, including pawnshop lending, may weaken such criminal responses (Karlan and Zinman 2010; Khanna et al. 2021; McCannon et al. 2022). In prewar Japan, pawnshops were used to respond to shocks and smooth consumption (Inoue 2021; Ogasawara 2024). By providing funds to households facing sudden shortages of living expenses or temporary income declines, pawnshop lending may reduce property-related crimes arising from short-term liquidity shortages.

The second channel is an increase in repayment burdens and debt stress. When repayment periods are short and borrowing is frequently repeated, credit may tighten rather than ease borrowers' cash flow. Studies of high-cost short-term credit indicate that access to credit may intensify repayment burdens and everyday stress rather than alleviate household hardship (Carrell and Zinman 2014; Melzer 2011). In addition, previous research has highlighted a relationship between debt problems and crime (Aaltonen et al. 2016). Pawnshop lending may therefore generate new pressure to secure repayment funds or create the risk of losing pledged assets, further increasing incentives for some types of property-related crime.

The third channel is the victimization of vulnerable people. Small loans that provide temporary cash or payment capacity to individuals who previously had little money may change not only borrowers' own behavior but also the opportunity structure facing offenders by creating new targets with accessible funds. Low-income individuals are more vulnerable to fraud (Lee and Soberon-ferrer 1997; Yu et al. 2022), and pawnshops are representative crime attractors where motivated offenders can more easily find targets that possess cash (Bernasco et al. 2013; Kubrin and Hipp 2016). Through this channel, pawnshop lending may increase property-related crimes committed by third parties rather than by borrowers themselves.

The fourth channel is the expansion of opportunities to convert stolen goods into cash. Pawnshops could function not only as credit providers but also as markets through which movable goods could be quickly monetized. The presence of pawnshops thickens stolen-goods markets, lowers the cost of monetizing stolen property, and increases the profitability of theft-related crime (d'Este 2020). Pawnshops in Japan had also been viewed as potential outlets for stolen goods and were subject to police surveillance (Shibuya et al. 1982). This channel connects pawnshop activity to theft through the market environment rather than borrower behavior.

Collectively, these four channels imply that pawnshop lending may reduce property-related crime through the first channel while increasing it through the second, third, and fourth channels. Because these mechanisms may operate simultaneously, and in opposite directions, the net relationship between pawnshop lending and property-related crime is not determined a priori. The following sections examine how institutional differences between public and private pawnshops may alter the relative importance of these channels and how the salience of each channel may vary across crime types.

### 3.2. Institutional Design and the Direction of Credit-Crime Relationships

Public and private pawnshops were distinct institutions in terms of interest rates, loan ceilings, actual loan sizes, and intended user groups. The question is which of the four crime-related channels these institutional differences were likely to activate more strongly and why public and private pawnshops may therefore exhibit different associations with property-related crime.

With low interest rates and loan ceilings, public pawnshops were likely most closely connected to the first channel. By functioning as a safety net for short-term livelihood needs, they may have reduced property-related crime arising from economic hardship. Simultaneously, these features, that channeled credit toward low-income groups, may also have led to the third channel. When economically vulnerable individuals gained access to small loans, they may have become more attractive targets for outside offenders, subsequently creating a countervailing force that increased some types of property-related crime. Public pawnshops should therefore not be understood as operating in only one direction.

Private pawnshops served a broader range of users and financial needs, including the acquisition of business funds, at higher interest rates. Their broader user base and higher interest burdens suggest a closer connection to both the first channel, in the form of liquidity provision extending beyond livelihood maintenance, and the second channel, namely repayment burdens and debt stress. Because such pawnshops also operated on a far larger scale and were privately owned, they may have been more strongly connected to the fourth channel than public pawnshops, as the lower perceived barrier to bringing in stolen goods may have thickened the market for monetizing stolen property.

In summary, while both institutional types could be linked to property-related crime through multiple channels, the combinations of channels most likely to operate differed between them. These differences inform the interpretation of the empirical results.

### 3.3. Crime-Type Heterogeneity within Property-Related Crime

The relationship between pawnshop credit and property-related crime may also vary depending on which crime type is examined. The three types analyzed in this study, theft, fraud/extortion, and embezzlement, differ in their required opportunity structures, the nature of expected returns, and their likely responsiveness to short-term financial shortages. This section considers which of the aforementioned channels are most likely to operate for each type of crime.

Theft is the type of property-related crime most directly connected to short-term credit. Theft allows offenders to obtain property within a short period under relatively low barriers to entry and is therefore likely to respond to sudden financial shortages or liquidity shocks. Thus, this crime is most naturally suppressed through the first channel and may also be induced through the fourth channel, given its close connection to markets for monetizing stolen goods. Although the third channel is not entirely irrelevant, pawnshops were present at extremely high density in Tokyo at the time, making it unlikely that their mere presence operated as a distinctive crime attractor for theft.

Fraud and extortion depend more strongly on interpersonal contact, deception, and the characteristics of victims, and they are strongly connected to the third victimization channel, in which crime increases as economically vulnerable people become potential targets.[9] The first channel is also relevant. If pawnshop lending reduces temporary financial needs, fraud and extortion may decline, although these offenses may be less responsive to short-term shortages than theft given the interpersonal engagement they require. The connection with the second channel, repayment burdens, is less robust, as whether the required repayment amount can be reliably secured through fraud or extortion is uncertain.

Embezzlement differs from the other two types of crime in that such crime involves diverting property the offender already possesses or manages on behalf of others, instead of acquiring property from outside. The expected return is therefore more predictable and typically larger than that of theft or fraud and extortion. When large financial shortages lie in the background, embezzlement may be substituted by pawnshop lending and decline through the first channel. Conversely, repayment burdens or cash-flow pressure may make embezzlement an attractive source of funds, thereby suggesting a relatively stronger connection with the second channel. Moreover, the relationship with the third and fourth channels is quite limited.

[9] Although fraud and extortion are combined in statistical data, this does not imply that they function identically. Nonetheless, unlike theft, both crimes rely heavily on connections with victims and personal opportunities.

The preceding discussion suggests that all three crime types may be related to the first channel, the relaxation of liquidity constraints, but that the most salient channels differ across crime types. Combining the two dimensions yields specific predictions. For public pawnshops, the liquidity relief channel implies a negative association with theft, the victimization channel implies a positive association with fraud and extortion, and all channels imply only a weak association with embezzlement. For private pawnshops, the predictions for theft and embezzlement are ambiguous because countervailing channels may offset one another. By contrast, for fraud and extortion, the liquidity relief channel is most salient and implies a negative association. Table 1 summarizes these conceptual expectations by pawnshop type and property-related crime type. To examine this heterogeneity, the relationship is empirically estimated separately for each crime type.

## 4. Methods

### 4.1. Data

This study uses a balanced ward-year panel for the 15 wards of the former Tokyo from 1927 to 1935. Although Tokyo expanded its municipal boundaries in 1932, the analysis is limited to the former area to maintain consistency over time. The main data consist of crime statistics, pawnshop statistics, and ward-level control variables. All monetary variables are deflated using the wholesale price index, with December 1929 as the base period.

The crime data are based on the Statistical Reports of the Tokyo Metropolitan Police Department. These statistics are available at the ward-year level. This study uses crime rates defined as the number of reported offenses per 1,000 populations. During the period examined, cleared cases are suspected to have been inflated, whereas the dark figure of crime is thought to have been relatively small because the police possessed extensive authority (Research and Training Institute, Ministry of Justice, ed., 1960). Reported offenses are therefore the most suitable measure of crime for the analysis. The baseline analysis first examines broad categories, namely total crime, aggregate property-related crime, and harm-related crime, before turning to more detailed property-related crime types, namely theft, fraud/extortion, and embezzlement. In this study, property-related crime is defined as the sum of these crimes. The theft measure includes pickpocketing, but excludes robbery, which also has a violent crime component. Harm-related crime is a category for which reporting is less likely to have been affected by legal revisions and, unlike property-related crime, is expected to have little direct relationship with pawnshop lending.[10]

---

[10] Microfinance reduces intimate partner violence but combining it with educational programs is important for this effect to occur (Matjasko et al., 2020). Thus, that historical pawnshops, which functioned only as financial institutions, reduced such violence is unlikely.

Because some crime classifications changed during the study period, the data are reaggregated to make them consistent with the 1927 classification and thereby allow time-series comparison. Specifically, fraud and extortion and theft and pickpocketing were each reported as combined types in 1927, whereas these offenses were reported separately from 1928 onward. Therefore, the relevant types in later years are summed to construct consistent series.[11]

The amounts and numbers of public pawnshop loans are based on the Annual Reports of the Tokyo City Social Welfare Bureau. As these data were originally recorded by branch and month, this study aggregates them to the ward-year level. The amounts and numbers of private pawnshop loans are based on the Statistical Yearbooks of Tokyo Prefecture.[12]

The redeemed and forfeited amounts used in the additional analyses are obtained from the Statistical Yearbooks of Tokyo City.[13] These amounts are annual observations recorded on a fiscal-year basis, and thus do not correspond exactly to calendar-year crime statistics. In addition, redemption or forfeiture sometimes occurred in the year following the loan. Hence, redeemed and forfeited amounts may reflect broader pawn transaction dynamics rather than the transaction outcomes of credit supplied in the same year.

[11] In practice, extortion and pickpocketing were only approximately ten and five percent as large as fraud and theft, respectively, so the combined series are driven mainly by fraud and theft.

[12] These variables are constructed following the procedures described in Inoue (2023).

[13] As these amounts of private pawnshops include those of public ones in the same ward, the private series is constructed by subtracting the public amounts from the ward-year total.

As control variables, this study uses the logarithm of population, male share of the population, income-taxpayer rate, individual business-taxpayer rate, social worker rate, and number of police officers per 1,000 population. These variables are included to capture differences in population size, population composition, economic conditions, social intervention, and the public-order environment. In particular, the individual business-taxpayer rate is used as a proxy for ward-level business conditions and commercial activity, while the income-taxpayer rate is interpreted as a measure of the size of high-income groups. The social worker rate is used to approximate both the extent of the local social response and the size of the poor population, while the police officer rate denotes police presence. All these variables are available at the ward-year level. Summary statistics of all variables are presented in Table 2.[14]

## 4.2. Empirical Strategy

The relationship between crime rates in ward $i$ in year $t$ and changes in public and private pawnshop lending is estimated using the following baseline specification:

$$CrimeRate_{it} = \beta_1 PublicPawn_{it} + \beta_2 PrivatePawn_{it} + \boldsymbol{X}'_{it}\boldsymbol{\gamma} + \nu_i + \mu_t + \delta_i t + \varepsilon_{it}.$$

---

[14] Detailed definitions and sources are provided in Appendix A.

Here, $CrimeRate_{it}$ is the crime rate per 1,000 population, $PublicPawn_{it}$ and $PrivatePawn_{it}$ are the amounts of public and private pawnshop lending, respectively, and $X_{it}$ is a vector of control variables. The specification includes ward fixed effects, $\nu_i$, year fixed effects, $\mu_t$, and ward-specific linear time trends, $\delta_i t$. Standard errors are clustered at the ward level. In the additional analyses, this specification is retained while the main independent variables are sequentially replaced with the number of loans, redeemed amounts, and forfeited amounts. These analyses are interpreted as comparisons that vary which aspect of pawnshop transactions is measured.

Because public pawnshop lending includes zero observations, the amounts of public and private pawnshop lending are transformed using the inverse hyperbolic sine.[15] The focus of this study is not the average amount borrowed per person but the scale of ward-level credit provision; that is, how much pawnshop credit was actually supplied in each ward. Thus, the main pawnshop variables are total amounts rather than per capita measures.[16]

[15] The inverse hyperbolic sine transformation is widely used as an alternative to the logarithmic transformation because it can reduce skewness in the distribution while retaining zero values. In particular, for sufficiently large values, its marginal effect approaches that of the logarithmic transformation.

[16] Average borrowing per individual may also be informative for considering the crime-related consequences of borrowing at the individual level, but this measure cannot be used because the population of pawnshop users is not observed. Dividing total lending by ward population may understate the scale of credit provision in wards where the utilization rate was low.

The amounts of public and private pawnshop loans are entered into the same regression as separate variables. Although both represent pawnshop credit, they differed in institutional design and intended user groups. Treating them separately allows the analysis to approach the mechanisms and institutional differences through which crime may be influenced more closely. In this specification, the relationship between within-ward changes in each type of pawnshop lending and crime rates is estimated while conditioning on the other type of pawnshop lending. Therefore, each coefficient indicates which crime type is associated with public and private credit provision, with changes in the other type of pawnshop credit held constant.

By including fixed effects, ward-specific linear trends, and control variables, this specification estimates the relationship between pawnshops and crime rates while accounting for many potential confounders. Ward fixed effects absorb characteristics specific to each ward that changed little over time, such as geographical location, while year fixed effects absorb annual shocks that affect all wards simultaneously, such as the Great Depression. Ward-specific linear trends further account for differences in medium- and long-run trends across wards. The control variables are included to capture time-varying factors.

Even with this specification, time-varying factors may remain. Reverse causality, in which changes in crime rates are related to the demand for or supply of pawnshop credit, cannot be ruled out. Thus, the estimates should not be interpreted as strict causal effects but as conditional associations between public and private credit provision and crime rates. Nevertheless, even under these limitations, comparing coefficient patterns across crime types is useful for examining the possibility that the relationship between pawnshop credit and property-related crime is not uniform, as well as what kinds of distinct associations are linked to the institutional differences between public and private pawnshops.

## 5. Results

### 5.1. Baseline Results

The relationship between public and private pawnshop loan amounts and broad crime categories, namely total, aggregate property-related, and harm-related crime, is presented in Table 3. No statistically clear association between either pawnshop variable and these broad crime categories is observed. This finding does not necessarily imply that pawnshop credit was unrelated to crime rates, as multiple channels may operate simultaneously in opposite directions, thereby offsetting one another in aggregate. The coefficients for harm-related crime are very small, which suggests that pawnshop credit was largely independent of violence.

In Table 4, clear heterogeneity is revealed when property-related crime is analyzed by type: theft, fraud/extortion, and embezzlement. Public pawnshop lending is significantly negatively associated with theft rates, significantly positively associated with fraud/extortion rates, and not significantly associated with embezzlement rates. In addition, the null hypothesis that the public pawnshop coefficients are equal across the three crime types is rejected. These patterns match the predictions in Table 1; the negative association with theft is consistent with the liquidity relief channel, and the positive association with fraud/extortion is consistent with the victimization channel. The absence of a clear relationship between public pawnshop loan amounts and embezzlement rates is also consistent with the hypothesis that embezzlement would show little response to the small, short-term credit that public pawnshops supplied.

The relationship between private pawnshops and property-related crime likewise becomes visible only when using crime outcomes by type. Private pawnshop lending is significantly negatively associated with fraud/extortion and embezzlement rates, but displays an insignificant association with theft rates. These results suggest that private pawnshop lending was more closely related to large or predictable needs than to small or urgent needs that may have led to theft. The null hypothesis that the private pawnshop coefficients are equal across crime types is not rejected, indicating that crime-type heterogeneity is not statistically confirmed as clearly for private pawnshops as for public ones.

Overall, the baseline results demonstrate that the relationship between pawnshop credit and property-related crime appears only when the latter is disaggregated, and that public and private pawnshops exhibit distinct associations probably because of differences in institutional design and user composition. The substantive magnitude and monetary implications of these heterogeneous associations are discussed later.[17]

## 5.2. Additional Analyses

In this section, additional analyses focusing on different aspects of pawnshop transactions are conducted to complement the interpretation of the baseline results. Loan counts instead of loan amounts are first used to examine whether the baseline findings remain consistent when pawnshop activity is measured by transaction frequency. Redeemed and forfeited amounts are then used to examine whether the relationship between pawnshop credit and crime rates is connected only to new lending or also to transaction stages involving repayment or collateral loss. Finally, private pawnshop loan counts are split according to average loan size to check whether the differences in results between public and private pawnshops can be explained simply by differences in loan size.

In Table 5, loan amounts are replaced with loan counts, which capture the frequency or breadth of pawnshop use rather than the total credit supplied. This specification examines whether the crime-type heterogeneity in the baseline results is driven by fluctuations in a small number of large transactions or reflects broader variation in use.

---

[17] Appendix B reports additional analyses and robustness checks for spatial spillovers, heterogeneity across economic characteristics, public-pawnshop presence, dynamic specifications and Granger-type tests, alternative crime measures and classifications, sample period, scaling of credit, inference, trend specification, and loan transformations. The main sign patterns are generally preserved.

The estimates closely replicate the baseline findings. Public pawnshop loan counts are significantly negatively associated with theft rates and significantly positively associated with fraud/extortion rates, while displaying no clear relationship with embezzlement rates. Therefore, the null hypothesis of equal coefficients across crime types is again rejected. Private pawnshop loan counts exhibit no significant relationship with theft but are significantly negatively associated with fraud/extortion and embezzlement. These results confirm that the crime-type heterogeneity reported in Table 4 is not specific to loan amounts.

In Table 6, loan amounts are replaced with redeemed amounts, thereby capturing the repayment-completion side of pawnshop transactions. Redeemed amounts reflect both the repayments actually borne by borrowers and the extent to which transactions were completed with collateral recovered.

For public pawnshops, the coefficient structure is highly similar to the baseline: a negative association with theft, a positive association with fraud/extortion, an insignificant association with embezzlement, and rejection of the null hypothesis of equal coefficients. For private pawnshops, the result for fraud/extortion remains largely unchanged from the baseline. However, the association with theft shifts to positive, and the negative association with embezzlement weakens. Since redemption necessarily follows lending, the similarity in the estimation results is reasonable given the strong correlation between redemption and loan amounts. Nonetheless, the results for theft and embezzlement change at the repayment stage, implying that while liquidity substitution may have reduced these crimes at the lending stage, the repayment pressure from higher-interest private loans may have partly offset the reduction.

In Table 7, loan amounts are replaced with forfeited amounts, thereby capturing transactions in which borrowers failed to repay their loans and to recover their collateral. Forfeited amounts are likely to reflect more severe cash-flow difficulties and repayment failure rather than the relaxation of liquidity constraints.

The estimates diverge considerably from the earlier results. Public pawnshop forfeited amounts display no clear relationship with any of the three crime types, and the null hypothesis of equal coefficients is not rejected. The results are consistent with the potential correspondence between theft and the liquidity relief channel and between fraud/extortion and the victimization channel, as both operate through credit provision rather than collateral loss.

For private pawnshops, forfeited amounts are significantly positively associated with theft, but display no clear relationship with fraud/extortion or embezzlement. These results suggest that cash-flow deterioration involving collateral loss may have been more associated with strong incentives for theft than for the other two crimes. Nevertheless, this interpretation warrants careful consideration as the potential for reverse causality offers an alternative explanation: an increase in theft may have led to an increase in pawned but unredeemed stolen goods.

Finally, whether the disparity in results between public and private pawnshops can be attributed to differences in the average loan size is examined in Table 8. The variable of private pawnshop loan counts is interacted with an indicator for whether the ward-year average loan amount per transaction is at or below, or above, 8.5 yen, which is the maximum observed average loan amount for public pawnshops. The estimated coefficients of private pawnshop loan counts are comparable between the two amount groups and replicate the results in Table 5. This suggests that larger transactions cannot solely explain the different associations with property-related crimes between public and private pawnshops. Broader institutional differences, including user composition, the nature of financial needs, or transaction behavior specific to the higher-interest private system, are likely to be involved.

The additional analyses largely corroborate the central implications of the baseline results and also offer some evidence regarding the relative plausibility of the channels discussed in the conceptual framework. The consistency of the public pawnshop findings across loan amounts, loan counts, and redeemed amounts, but not forfeited amounts, supports the liquidity relief and victimization channels, both of which operate at the point of credit provision. For private pawnshops, the stability of the negative associations with fraud/extortion and embezzlement across multiple measures strengthens the case for the relaxation of liquidity constraints as a primary mechanism. By contrast, the debt burden and stolen-goods market channels remain ambiguous, even though the positive association between private pawnshop forfeitures and theft provides suggestive evidence for the former. Furthermore, these analyses also indicate that the relationship between pawnshop credit and property-related crime varies not only by pawnshop and crime type but also by transaction stage.

## 6. Discussion

According to the baseline results, the relationship between pawnshop lending and property-related crime differed substantially across both crime types and institutional forms. To assess the substantive scale of these associations, the baseline estimates are translated into implied changes in recorded offenses and monetized victim losses.

For each ward $i$ and year $t$ over 1928–1935, using 1927 as the baseline year, the change in pawnshop lending relative to 1927 for pawnshop type $s$ is first calculated:

$$\Delta \mathrm{IHS}(Loan_{s,i,t}) = \mathrm{IHS}\big(Loan_{s,i,t}\big) - \mathrm{IHS}\big(Loan_{s,i,1927}\big).$$

The implied change in recorded offenses is then calculated as:

$$\Delta Cases_{c,s,i,t} = \beta_{c,s} \times \Delta \mathrm{IHS}(Loan_{s,i,t}) \times \frac{pop_{i,t}}{1000},$$

where $c$ indexes crime type, namely theft, fraud/extortion, or embezzlement. The coefficient $\beta_{c,s}$ is taken from the baseline regressions reported in Table 4. The aggregate implied change in recorded offenses is obtained by summing this value across wards and years. To monetize these changes, implied damage averted at the ward-year level is computed as:

$$Damage_averted_{c,s,i,t} = -\Delta Cases_{c,s,i,t} \times damage_per_case_{c,t}.$$

The damage-per-case measure for each property-related crime type is calculated by dividing the annual total damages in 15 wards, sourced from the Statistical Yearbooks of Tokyo City, by the number of crimes. This measure is deflated to December 1929 prices. Aggregate values are obtained by summing ward-year damage averted across 1928–1935; positive and negative values indicate damage avoided and additional damage, respectively. As the calculation uses ward-year damages, the aggregate monetized values do not mechanically equal aggregate implied changes in recorded offenses multiplied by average damage per offense.

The results for public pawnshops are reported in Panel A of Table 9. Public pawnshop lending is associated with approximately 5,470 fewer theft offenses over 1928–1935, equivalent to 1.11 percent of recorded theft offenses during the period. In monetary terms, this corresponds to 185.44 thousand yen in implied damage avoided from theft. However, public pawnshop lending is also associated with approximately 2,675 additional fraud/extortion offenses, or 4.09 percent of recorded offenses of that type. As fraud and extortion involved much larger losses per offense than theft, this corresponds to 646.75 thousand yen in implied additional damage. Aggregating theft and fraud/extortion, public pawnshop lending is associated with 0.50 percent fewer recorded offenses overall but with 461.30 thousand yen in net additional damage.

The corresponding results for private pawnshops are reported in Panel B. Private pawnshop lending is associated with reductions equivalent to 6.42 and 8.76 percent of recorded fraud/extortion and embezzlement offenses during 1928–1935, respectively, and with implied damage avoided of 1,140.99 and 1,062.40 thousand yen, respectively. Aggregating these two types, private pawnshop lending is associated with 7.19 percent fewer recorded offenses and 2,203.39 thousand yen in implied damage avoided.

Overall, these calculations show that crime-type heterogeneity matters not only statistically but also economically. Public pawnshops are associated with fewer recorded property-related offenses overall because the reduction in theft outweighs the increase in fraud/extortion in offense counts. However, once crimes are weighted by victim losses, fraud/extortion dominates, thus producing net additional damage. By contrast, private pawnshops are negatively associated with higher-loss crime types, thereby generating substantial implied reductions in both offense counts and monetized losses. These findings reinforce the central argument that the social implications of pawnshop credit depended jointly on institutional design and crime type, and that evaluations based solely on aggregate property-related-crime rates risk obscuring important differences in both the composition and the economic consequences of crime.

## 7. Conclusions

By distinguishing both types of pawnshop institutions and property-related crime, this study explored the relationship between pawnshop credit and property-related crime in interwar Tokyo. Ward-level panel data for 15 wards of Tokyo from 1927 to 1935 reveal the heterogeneous associations: public pawnshop lending was negatively and positively associated with theft and fraud/extortion, respectively, while private pawnshop lending was negatively associated with fraud/extortion and embezzlement. These results indicate that pawnshops, as a form of fringe banking, were related to property-related crime beyond theft, and that the relationships cannot be understood as uniformly crime-reducing or crime-increasing.

The findings also demonstrate why institutional design matters. With low interest rates, small loans, and a low-income target group, public pawnshops appear to have been linked both to liquidity relief and to the formation of vulnerable targets. Private pawnshops, which served a broader set of financial needs, were associated with reductions in crime types involving larger victim losses. The back-of-the-envelope monetization exercise reinforces this point: conclusions could differ depending on whether recorded offense counts or victim losses are evaluated. However, in either case, private pawnshops may have contributed more to public order than public pawnshops in interwar Tokyo.

Even with the limited identification of causal effects, the evidence suggests that small-scale credit institutions may have played a role in shaping urban order through channels that depended on crime type. More broadly, the findings suggest that the history of popular credit should be integrated more closely with the history of urban governance. The Tokyo case suggests that public pawnshops can be regarded not only as welfare institutions but also as urban policy instruments located at the intersection of poor relief and public order. Moreover, private pawnshops may have served as urban credit infrastructure closely connected to the maintenance of public order.

## Acknowledgments

I would like to express my gratitude to participants in the CIGS Meeting. This work was supported by Grant-in-Aid for Early-Career Scientists (25K16696). All errors are my own.

References

Aaltonen, Mikko, Atte Oksanen, and Janne Kivivuori. 2016. "Debt Problems and Crime." *Criminology* 54 (2): 307–31.

Bank of Japan. 1986. *Nihon Ginko hyakunenshi, shiryo-hen [One Hundred-Year History of the Bank of Japan, supplementary materials]*. Tokyo: Bank of Japan.

Barth, James R., Jitka Hilliard, John S. Jahera, Kang B. Lee, and Yanfei Sun. 2020. "Payday Lending, Crime, and Bankruptcy: Is There a Connection?" *The Journal of Consumer Affairs* 54 (4): 1159–77.

Becker, Gary S. 1968. "Crime and Punishment: An Economic Approach." *Journal of Political Economy* 76 (2): 169–217.

Bernasco, Wim, Richard Block, and Stijn Ruiter. 2013. "Go Where the Money Is: Modeling Street Robbers' Location Choices." *Journal of Economic Geography* 13 (1): 119–43.

Blakeslee, David S., and Ram Fishman. 2018. "Weather Shocks, Agriculture, and Crime: Evidence from India." *The Journal of Human Resources* 53 (3): 750–82.

Britto, Diogo G. C., Paolo Pinotti, and Breno Sampaio. 2022. "The Effect of Job Loss and Unemployment Insurance on Crime in Brazil." *Econometrica: Journal of the Econometric Society* 90 (4): 1393–1423.

Cabinet Statistics Bureau. 1933. *Kokusei chosa hokoku, Showa 5 nen, dai 4-kan, fuken-hen: Tokyo-fu [Population Census Report, 1930, volume 4, Prefectural section: Tokyo Prefecture]*. Tokyo: Cabinet Statistics Bureau.

Carbonell-Esteller, Montserrat. 2012. "Montes de Piedad and Savings Banks as Microfinance Institutions on the Periphery of the Financial System of Mid-Nineteenth-Century Barcelona." *Business History* 54 (3): 363–80.

Carboni, Mauro, and Massimo Fornasari. 2019. "The 'Untimely' Demise of a Successful Institution: The Italian Monti Di Pietà in the Nineteenth Century." *Financial History Review* 26 (2): 147–70.

Carrell, Scott, and Jonathan Zinman. 2014. "In Harm's Way? Payday Loan Access and Military Personnel Performance." *Review of Financial Studies* 27 (9): 2805–2840.

Carter, Susan Payne, Kuan Liu, Paige Marta Skiba, and Justin Sydnor. 2022. "Time to Repay or Time to Delay? The Effect of Having More Time before a Payday Loan Is Due." *American Economic Journal. Applied Economics* 14 (4): 91–126.

Caskey, John P. 1991. "Pawnbroking in America: The Economics of a Forgotten Credit Market." *Journal of Money, Credit, and Banking* 23 (1): 85.

d'Este, Rocco. 2020. "The Effects of Stolen-Goods Markets on Crime: Pawnshops, Property Theft, and the Gold Rush of the 2000s." *The Journal of Law & Economics* 63 (3): 449–72.

Ehrlich, Isaac. 1973. "Participation in Illegitimate Activities: A Theoretical and Empirical Investigation." *Journal of Political Economy* 81 (3): 521–565.

Field, Erica, and Rohini Pande. 2008. "Repayment Frequency and Default in Microfinance: Evidence from India." *Journal of the European Economic Association* 6 (2–3): 501–9.

Field, Erica, Rohini Pande, John Papp, and Natalia Rigol. 2013. "Does the Classic Microfinance Model Discourage Entrepreneurship among the Poor? Experimental Evidence from India." *American Economic Review* 103 (6): 2196–2226.

Giné, Xavier, and Dean S. Karlan. 2014. "Group versus Individual Liability: Short- and Long-Term Evidence from Philippine Microcredit Lending Groups." *Journal of Development Economics* 107 (March): 65–83.

Inoue, Tatsuki. 2021. "The Role of Pawnshops in Risk Coping in Early Twentieth-Century Japan." *Financial History Review* 28 (3): 319–43.

Inoue, Tatsuki. 2023. "Health impacts of public pawnshops in industrializing Tokyo." *arXiv preprint arXiv:2305.09352*.

Ioku, Shigehiko, and Masato Shizume. 2014. "Kindai nihon no shomin kinyu: Tokyo-shi Shiba-ku T shichiten no kenkyu" [Finance for common people in modern Japan: the case of "T" pawnshop in Shiba ward in Tokyo]. *Shakai Keizaishi Gaku*, 80 (3): 3–8.

Karlan, Dean S., and Jonathan Zinman. 2010. "Expanding Credit Access: Using Randomized Supply Decisions to Estimate the Impacts." *Review of Financial Studies* 23 (1): 433–64.

Khanna, Gaurav, Carlos Medina, Anant Nyshadham, Christian Posso, and Jorge Tamayo. 2021. "Job Loss, Credit, and Crime in Colombia." *American Economic Review. Insights* 3 (1): 97–114.

Koukei, Hisamatsu. 1928. "Kōsetsu shichigyō no shakaijō ni oyobosu eikyō" [The social effects of public pawnbroking]. *Tōkyō-fu Shakai Jigyō Kyōkai Kaihō* 41: 24–5.

Kubrin, Charis E., and John R. Hipp. 2016. "Do Fringe Banks Create Fringe Neighborhoods? Examining the Spatial Relationship between Fringe Banking and Neighborhood Crime Rates." *Justice Quarterly* 33 (5): 755–84.

Kubrin, Charis E., Gregory D. Squires, Steven M. Graves, and Graham C. Ousey. 2011. "Does Fringe Banking Exacerbate Neighborhood Crime Rates? Investigating the Social Ecology of Payday Lending." *Criminology & Public Policy* 10 (2): 437–66.

Lee, Anne M., Randy Gainey, and Ruth Triplett. 2014. "Banking Options and Neighborhood Crime: Does Fringe Banking Increase Neighborhood Crime?" *American Journal of Criminal Justice* 39 (3): 549–70.

Lee, Jinkook, and Horacio Soberon-ferrer. 1997. "Consumer Vulnerability to Fraud: Influencing Factors." *The Journal of Consumer Affairs* 31 (1): 70–89.

Matjasko, Jennifer L., Ashley Schappell D'Inverno, Khiya J. Marshall, and Megan C. Kearns. 2020. "Microfinance and Violence Prevention: A Review of the Evidence and Adaptations for Implementation in the U.S." *Preventive Medicine* 133 (106017): 106017.

McCannon, Bryan C., Zachary Porreca, and Zachary Rodriguez. 2022. "Three Golden Balls: Pawn Shops and Crime." *SSRN Electronic Journal*. https://doi.org/10.2139/ssrn.4119571.

McLaughlin, Eoin. 2013. "An Experiment in Banking the Poor: The Irish Mont-de-Piété, c. 1830–1850." *Financial History Review* 20 (1): 49–72.

McLaughlin, Eoin, and Rowena Pecchenino. 2022. "Fringe Banking and Financialization: Pawnbroking in Pre-famine and Famine Ireland." *The Economic History Review* 75 (3): 903–31.

Melzer, Brian T. 2011. "The Real Costs of Credit Access: Evidence from the Payday Lending Market." *The Quarterly Journal of Economics* 126 (1): 517–55.

Morse, Adair. 2011. "Payday Lenders: Heroes or Villains?" *Journal of Financial Economics* 102 (1): 28–44.

Murhem, Sofia. 2016. "Credit for the Poor: The Decline of Pawnbroking 1880–1930." *European Review of Economic History* 20 (2): 198–214.

Murhem, Sofia, and Göran Ulväng. 2023. "Pawning and Pawners in the Industrial Era: Evidence from Sweden, 1870 to 1950." *History of Retailing and Consumption* 9 (1): 38–56.

Nakagawa, Kiyoshi. 1985. *Nihon no toshi kaso [The urban lower classes in Japan]*. Tokyo: Keiso Shobo.

Obinata, Sumio. 1993. *Keisatsu no shakaishi [A social history of the police]*. Tokyo: Iwanami Shoten.

Ogasawara, Kota. 2024. "Consumption Smoothing in the Working-Class Households of Interwar Japan." *The Journal of Economic History* 84 (1): 111–48.

Pastureau, Guillaume, and Bertrand Blancheton. 2014. "The Success of the Bordeaux Mont-de-Piété, 1801–1913." *Financial History Review* 21 (3): 281–99.

Research and Training Institute of the Ministry of Justice, ed. (1960). *Hanzai hakusho, Showa 35 nenban: Waga kuni ni okeru hanzai to sono taisaku [White Paper on Crime, 1960: Crime and crime control in Japan]*. Tokyo: Printing Bureau, Ministry of Finance.

Research and Training Institute of the Ministry of Justice, ed. (1968). *Hanzai hakusho, Showa 43 nenban: Hanzai to hanzaisha no shogu, sono genkyo to 100 nenkan no suii [White Paper on Crime, 1968: Crime and the treatment of offenders, current conditions and trends over one hundred years]*. Tokyo: Printing Bureau, Ministry of Finance.

Shackman, Joshua D., and Glen Tenney. 2006. "The Effects of Government Regulations on the Supply of Pawn Loans: Evidence from 51 Jurisdictions in the U.S." *Journal of Financial Services Research* 30 (1): 69–91.

Shibuya, Ryuichi, Suzuki, Kameji, Ishiyama, Shojiro. 1982. *Nihon no shichiya [Pawnshops in Japan]*. Tokyo: Waeda Daigaku Shuppambu.

Social Welfare Bureau. 1926. *Koeki shichiko no genzei [Current conditions of public pawnshops]*. Tokyo.

Stegman, Michael A. 2007. "Payday Lending." *Journal of Economic Perspectives* 21 (1): 169–90.

Tokyo City Office. 1926. *Shichiyagyo no tokei chosa [Statistical survey on pawnbroking]*. Tokyo.

Tokyo City Office. 1931. *Musansha kinyu kikan to shite no shichiya no kenkyu [A study of pawnshops as financial institutions for the propertyless]*. Tokyo.

Tokyo City Social Welfare Bureau. 1926. *Tokyo shinai oyobi gunbu ni okeru shichiya ni kansuru chosa [Survey on pawnshops in Tokyo City and the county districts]*. Tokyo: Tokyo City Social Welfare Bureau, Social Department.

Tokyo City Social Welfare Bureau. 1933. *Tokyoshi yohogosha chosa [Survey of protection-required people in Tokyo City]*. Tokyo.

Tokyo Institute for Municipal Research. 1925. *Toshi kinyu no gaikyo [Overview of urban finance]*. Tokyo.

Tokyo Metropolitan Police Department. 1934. *Keishicho tokei ippan, Showa 8 nen [Statistical overview of the Tokyo Metropolitan Police Department, 1933]*. Tokyo: Statistics Section, Documents Division, Commissioner-General's Secretariat.

Tokyo Prefecture Department of Academic Affairs. 1935. *Saimin kinyu ni kansuru chosa [Survey on finance among the poor]*. Tokyo.

Woloson, Wendy A. 2009. *In Hock: Pawning in America from Independence through the Great Depression*. Chicago: University of Chicago Press.

Yu, Lei, Gary Mottola, Lisa L. Barnes, Olivia Valdes, Robert S. Wilson, David A. Bennett, and Patricia A. Boyle. 2022. "Financial Fragility and Scam Susceptibility in Community Dwelling Older Adults." *Journal of Elder Abuse & Neglect* 34 (2): 93–108.

Data Sources

Statistical Division of the Commerce and Industry Ministry. (1929, 1931, 1933–1936). *Oroshiuri bukka tokeihyo [Statistics of wholesale price index, 1926–1935 editions]*. Tokyo.

Tokyo City Social Welfare Bureau. 1928–1935. *Tokyoshi Shakaikyoku nenpo, dai 8–15 kai [Annual reports of the Tokyo City Social Welfare Bureau, Volumes 8–15]*. Tokyo.

Tokyo City Office. 1929–1937. *Tokyoshi tokei nenpyo, dai 25–33 kai [Statistical Yearbooks of Tokyo City, Volumes 25–33]*. Tokyo.

Tokyo Metropolitan Police Department. 1921–1927. *Keishicho tokeisho [Statistical Reports of the Tokyo Metropolitan Police Department, 1920–1926 editions]*. Tokyo.

Tokyo Metropolitan Police Department. 1928–1939, 1941–1942. *Keishicho tokeisho, dai 37–50 kai [Statistical Reports of the Tokyo Metropolitan Police Department, Volumes 37–50]*. Tokyo.

Tokyo Prefecture. 1926–1937. *Tokyofu tokeisho [Statistical Yearbooks of Tokyo Prefecture, 1924–1935 editions]*. Tokyo.

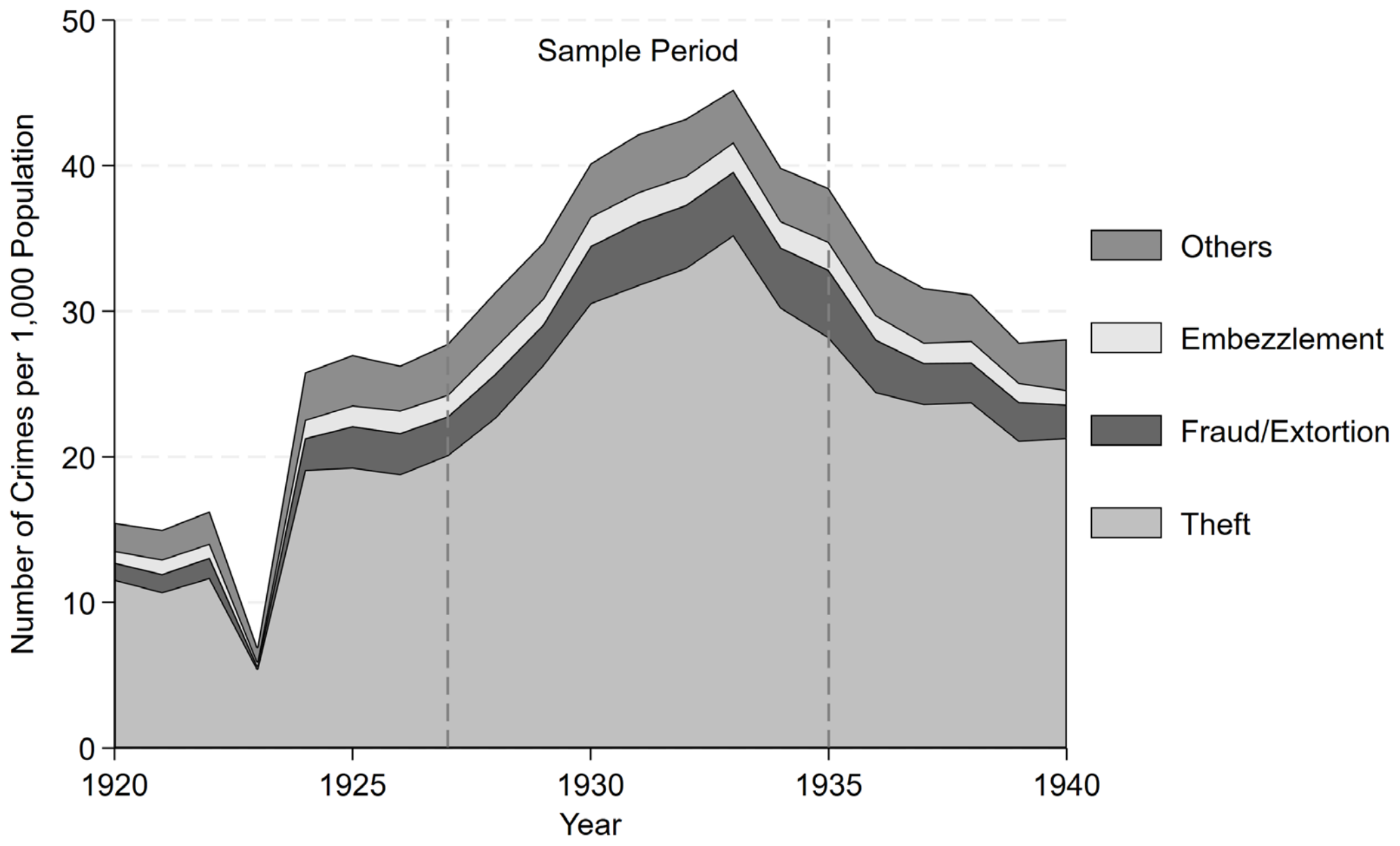

Figure 1: Crime Rates in Tokyo

Notes: The figure reports annual crime rates for the former 15 wards of Tokyo from 1920 to 1940. Crime rates are measured as reported offenses per 1,000 populations.

Sources: Tokyo Metropolitan Police Department (1921–1927); Tokyo Metropolitan Police Department (1928–1939, 1941–1942).

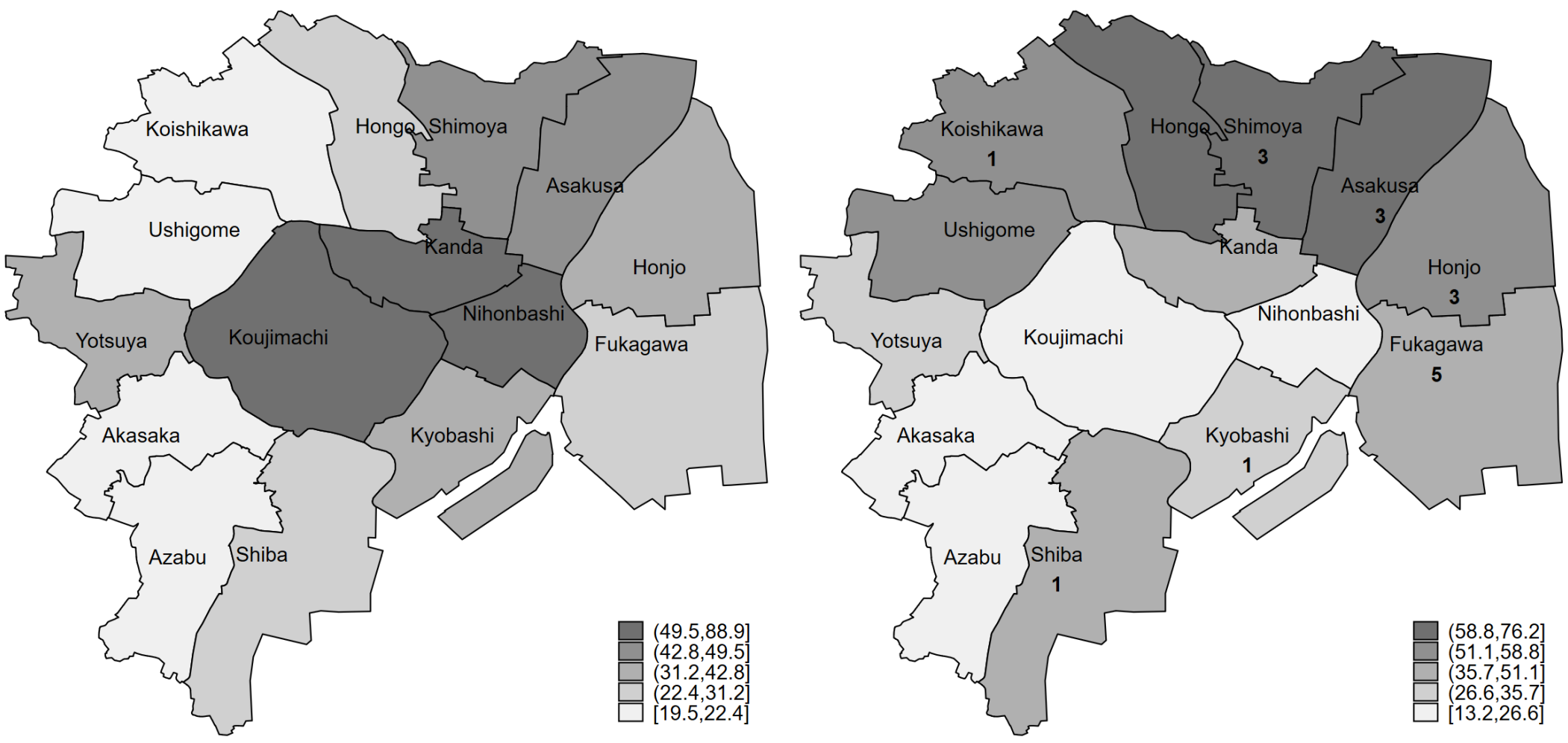

(a) Crime rate

(b) Pawnshops

Figure 2: Spatial Distribution of Crime Rates and Pawnshops

Notes: Panel A reports the average crime rate from 1927 to 1935, measured as reported offenses per 1,000 populations. In Panel B, ward shading reports the average number of private pawnshops from 1927 to 1935, and ward labels report the number of public pawnshops in 1935. The number of public pawnshops is unchanged from 1930 onward.

Sources: Tokyo Metropolitan Police Department (1928–1939, 1941–1942, vol. 37–45); Tokyo City Social Welfare Bureau (1928–1935); Tokyo Prefecture (1926–1937).

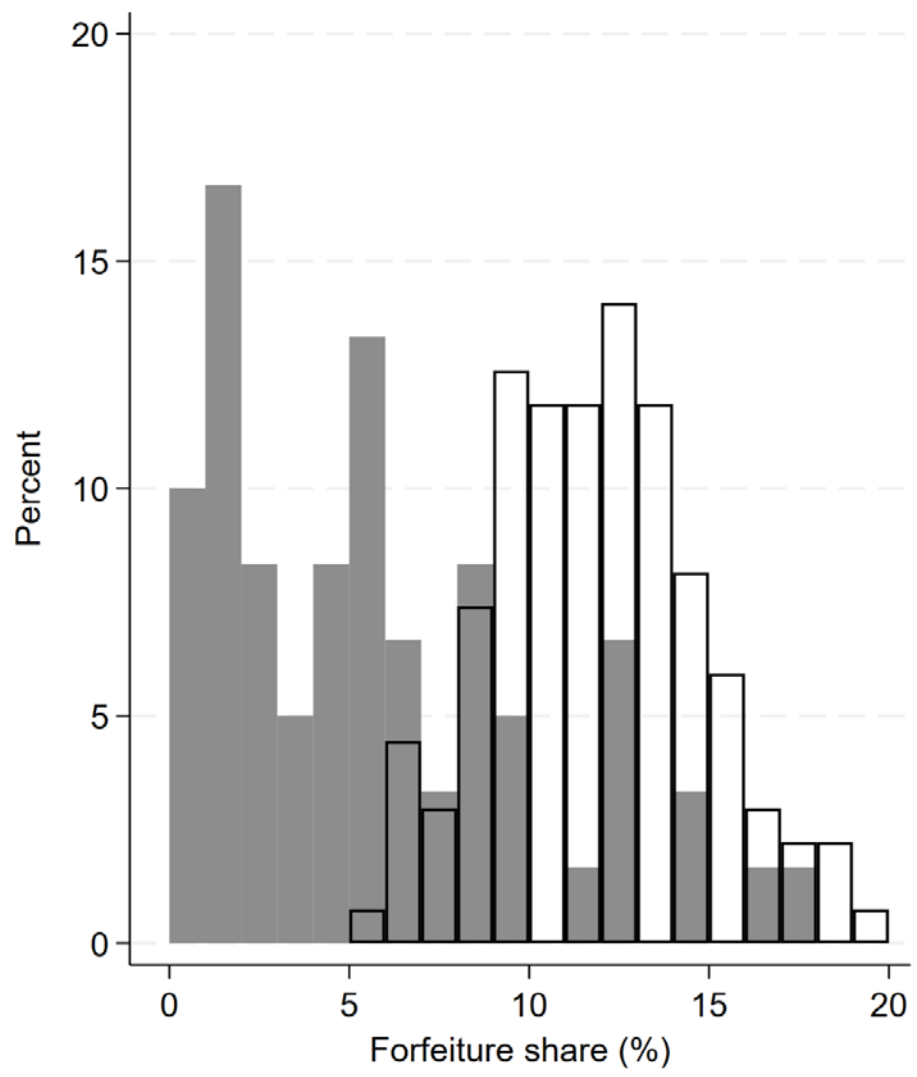

(a) Forfeiture Share

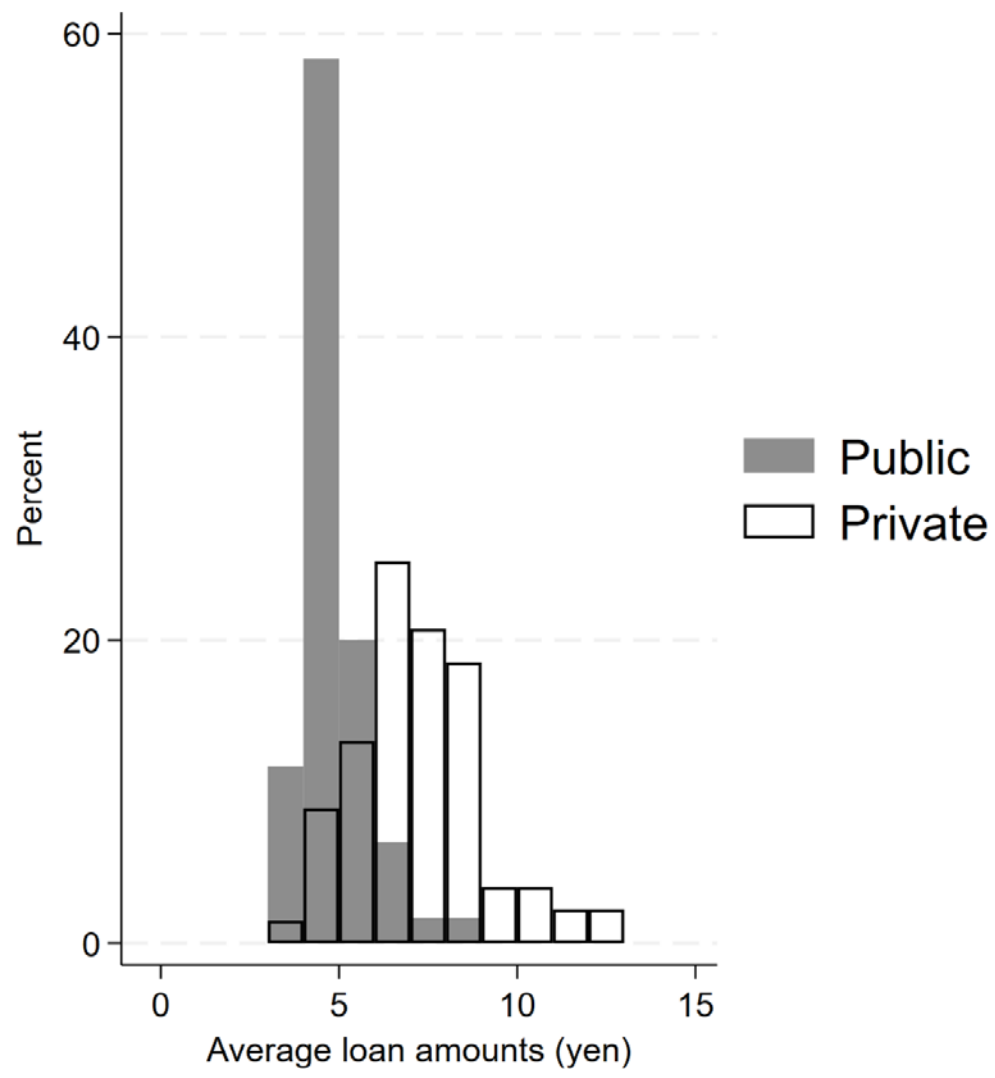

(b) Average loan amounts

Figure 3: Differences between Public and Private Pawnshops in 1927–1935

Notes: Panel A shows the forfeiture rate among completed pawnshop transactions at the ward-year level from 1927 to 1935, calculated on a transaction-count basis. Panel B shows the average loan amount per transaction at the ward-year level. Monetary values are real values adjusted by the wholesale price index, with December 1929 as the base period.

Sources: Tokyo City Social Welfare Bureau (1928–1935); Tokyo City Office (1929–1937); Tokyo Prefecture (1926–1937); Statistical Division of the Commerce and Industry Ministry (1929, 1931, 1933–1936).

Table 1: Conceptual Summary of Likely Channels

Panel A. Public Pawnshops

| Property-related crime type | LR | DB | VT | SG |
|---|---|---|---|---|
| Theft | ● | | ○ | |
| Fraud/Extortion | ○ | | ● | |
| Embezzlement | | | | |

Panel B. Private Pawnshops

| Property-related crime type | LR | DB | VT | SG |
|---|---|---|---|---|
| Theft | ○ | ○ | | ○ |
| Fraud/Extortion | ● | ○ | | |
| Embezzlement | ● | ● | | |

Notes: LR = liquidity relief; DB = debt burden; VT = victim targeting; SG = stolen-goods market. ● indicates the channel most likely to be salient. ○ indicates a secondary or potentially relevant channel. This table summarizes the conceptual expectations and does not imply that each channel is separately identified.

Table 2: Summary Statistics

| | Mean | SD | Min | Max | Obs. |
|---|---|---|---|---|---|
| Crime Rates (per 1,000 population) | | | | | |
| All crimes | 39.17 | 19.47 | 12.95 | 98.64 | 135 |
| Property-Related crimes | 35.42 | 18.45 | 11.11 | 90.73 | 135 |
| Harm-Related crimes | 0.47 | 0.27 | 0.11 | 1.77 | 135 |
| Theft | 29.45 | 16.68 | 8.37 | 80.00 | 135 |
| Fraud/Extortion | 3.98 | 1.71 | 1.44 | 13.11 | 135 |
| Embezzlement | 1.99 | 0.69 | 0.63 | 3.73 | 135 |
| | | | | | |
| Pawnshop Variables | | | | | |
| Loan amount (1,000 yen) | | | | | |
| Public | 58.94 | 85.02 | 0.00 | 323.75 | 135 |
| Private | 1,483.08 | 687.50 | 469.82 | 3,444.13 | 135 |
| Number of loans (thousand) | | | | | |
| Public | 12.63 | 19.10 | 0.00 | 84.00 | 135 |
| Private | 223.45 | 121.85 | 53.37 | 607.82 | 135 |
| Redemption amount (1,000 yen) | | | | | |
| Public | 59.86 | 87.01 | 0.00 | 338.89 | 135 |
| Private | 1282.87 | 566.83 | 410.82 | 2974.10 | 135 |
| Forfeiture amount (1,000 yen) | | | | | |
| Public | 3.98 | 8.15 | 0.00 | 52.85 | 135 |
| Private | 159.82 | 98.26 | 28.64 | 513.64 | 135 |
| | | | | | |
| Control Variables | | | | | |
| Population (1,000 persons) | 136.65 | 57.84 | 49.44 | 272.04 | 135 |
| Male share (%) | 53.39 | 2.91 | 47.69 | 60.36 | 135 |
| Business taxpayers (% of pop.) | 2.87 | 1.02 | 1.45 | 6.34 | 135 |
| Income taxpayers (% of pop.) | 3.92 | 1.44 | 1.32 | 6.69 | 135 |
| Social workers (% of pop.) | 0.03 | 0.02 | 0.00 | 0.07 | 135 |
| Police officers (per 1,000 pop.) | 3.69 | 1.67 | 2.10 | 9.84 | 135 |

Notes: Crime rates are defined as the number of reported offenses per 1,000 populations. Property-related crime is defined as the sum of theft, fraud and extortion, and embezzlement. Theft includes pickpocketing but excludes robbery. Harm-related crime is defined as the sum of harm, injury resulting in death, and simple assault. Public and private loan amounts, redemption amounts, and forfeiture amounts are reported in thousands of yen. Monetary values are real values adjusted by the wholesale price index, with December 1929 as the base period. Loan counts are reported in thousands. The male share, business-taxpayer rate, income-taxpayer rate, and social worker rate are measured as percentages of ward population. Police officers are measured per 1,000 population.
Sources: Tokyo Metropolitan Police Department (1928–1939, 1941–1942, vol. 37–45); Tokyo City Social Welfare Bureau (1928–1935); Tokyo City Office (1929–1937); Tokyo Prefecture (1926–1937); Statistical Division of the Commerce and Industry Ministry (1929, 1931, 1933–1936).

Table 3: Relationship Between Pawnshop Loans and Broad Crime Categories

| | All crimes | Property-related crimes | Harm-related crimes |
|---|---|---|---|
| | (1) | (2) | (3) |
| Public pawn loan amount | -0.178 | -0.151 | -0.003 |
| | (0.163) | (0.145) | (0.006) |
| Private pawn loan amount | -2.696 | -2.897 | -0.018 |
| | (4.753) | (4.706) | (0.194) |
| All control variables | Yes | Yes | Yes |
| Ward FE | Yes | Yes | Yes |
| Year FE | Yes | Yes | Yes |
| Time trend | Yes | Yes | Yes |
| Observations | 135 | 135 | 135 |

***, **, and * represent statistical significance at the 1%, 5%, and 10% levels, respectively.

Notes: The dependent variables are crime rates per 1,000 populations. Public and private pawn loan amounts are transformed using the inverse hyperbolic sine. Monetary values are real values adjusted by the wholesale price index, with December 1929 as the base period. All specifications include the full set of control variables, ward fixed effects, year fixed effects, and ward-specific linear trends. Control variables are the logarithm of population, male share of the population, business-taxpayer rate, income-taxpayer rate, social worker rate, and police officers per 1,000 populations. Standard errors clustered at the ward level are reported in parentheses.

Table 4: Relationship Between Pawnshop Loans and Property-Related Crime Types

| | Theft | Fraud/Extortion | Embezzlement |
|---|---|---|---|
| | (1) | (2) | (3) |
| Public pawn loan amount | -0.276* | 0.135*** | -0.009 |
| | (0.139) | (0.034) | (0.019) |
| Private pawn loan amount | -0.487 | -1.443** | -0.967*** |
| | (4.837) | (0.613) | (0.241) |
| All control variables | Yes | Yes | Yes |
| Ward FE | Yes | Yes | Yes |
| Year FE | Yes | Yes | Yes |
| Time trend | Yes | Yes | Yes |
| Observations | 135 | 135 | 135 |
| Public coefficients equal (p-value) | | 0.011 | |
| Private coefficients equal (p-value) | | 0.834 | |

***, **, and * represent statistical significance at the 1%, 5%, and 10% levels, respectively.

Notes: The dependent variables are crime rates per 1,000 populations. Public and private pawn loan amounts are transformed using the inverse hyperbolic sine. Monetary values are real values adjusted by the wholesale price index, with December 1929 as the base period. All specifications include the full set of control variables, ward fixed effects, year fixed effects, and ward-specific linear trends. Control variables are the logarithm of population, male share of the population, business-taxpayer rate, income-taxpayer rate, social worker rate, and police officers per 1,000 populations. Standard errors clustered at the ward level are reported in parentheses. The reported p-values test the null hypothesis that the pawnshop coefficients are equal across theft, fraud and extortion, and embezzlement.

Table 5: Relationship Between Loan Counts and Property-Related Crime Types

| | Theft | Fraud/Extortion | Embezzlement |
|---|---|---|---|
| | (1) | (2) | (3) |
| Public pawn loan count | -0.358** | 0.145*** | -0.023 |
| | (0.139) | (0.043) | (0.022) |
| Private pawn loan count | -3.701 | -1.056* | -0.724** |
| | (3.674) | (0.544) | (0.269) |
| All control variables | Yes | Yes | Yes |
| Ward FE | Yes | Yes | Yes |
| Year FE | Yes | Yes | Yes |
| Time trend | Yes | Yes | Yes |
| Observations | 135 | 135 | 135 |
| Public coefficients equal (p-value) | | 0.011 | |
| Private coefficients equal (p-value) | | 0.592 | |

***, **, and * represent statistical significance at the 1%, 5%, and 10% levels, respectively.

Notes: The dependent variables are crime rates per 1,000 populations. Public and private pawn loan counts are transformed using the inverse hyperbolic sine. All specifications include the full set of control variables, ward fixed effects, year fixed effects, and ward-specific linear trends. Control variables are the logarithm of population, male share of the population, business-taxpayer rate, income-taxpayer rate, social worker rate, and police officers per 1,000 populations. Standard errors clustered at the ward level are reported in parentheses. The reported p-values test the null hypothesis that the pawnshop coefficients are equal across theft, fraud and extortion, and embezzlement.

Table 6: Relationship Between Redemption Amounts and Property-Related Crime Types

| | Theft | Fraud/Extortion | Embezzlement |
|---|---|---|---|
| | (1) | (2) | (3) |
| Public pawn redemption amount | -0.315** | 0.129*** | -0.021 |
| | (0.133) | (0.036) | (0.020) |
| Private pawn redemption amount | 3.491 | -2.025* | -0.656 |
| | (3.262) | (1.030) | (0.493) |
| All control variables | Yes | Yes | Yes |
| Ward FE | Yes | Yes | Yes |
| Year FE | Yes | Yes | Yes |
| Time trend | Yes | Yes | Yes |
| Observations | 135 | 135 | 135 |
| Public coefficients equal (p-value) | | 0.008 | |
| Private coefficients equal (p-value) | | 0.275 | |

***, **, and * represent statistical significance at the 1%, 5%, and 10% levels, respectively.

Notes: The dependent variables are crime rates per 1,000 populations. Public and private pawn redemption amounts are transformed using the inverse hyperbolic sine. Monetary values are real values adjusted by the wholesale price index, with December 1929 as the base period. All specifications include the full set of control variables, ward fixed effects, year fixed effects, and ward-specific linear trends. Control variables are the logarithm of population, male share of the population, business-taxpayer rate, income-taxpayer rate, social worker rate, and police officers per 1,000 populations. Standard errors clustered at the ward level are reported in parentheses. The reported p-values test the null hypothesis that the pawnshop coefficients are equal across theft, fraud and extortion, and embezzlement.

Table 7: Relationship Between Forfeiture Amounts and Property-Related Crime Types

| | Theft | Fraud/Extortion | Embezzlement |
|---|---|---|---|
| | (1) | (2) | (3) |
| Public pawn forfeiture amount | 0.114 | 0.008 | 0.001 |
| | (0.299) | (0.067) | (0.032) |
| Private pawn forfeiture amount | 3.611** | -1.027 | -0.067 |
| | (1.663) | (0.836) | (0.270) |
| All control variables | Yes | Yes | Yes |
| Ward FE | Yes | Yes | Yes |
| Year FE | Yes | Yes | Yes |
| Time trend | Yes | Yes | Yes |
| Observations | 135 | 135 | 135 |
| Public coefficients equal (p-value) | | 0.931 | |
| Private coefficients equal (p-value) | | 0.028 | |

***, **, and * represent statistical significance at the 1%, 5%, and 10% levels, respectively.
Notes: The dependent variables are crime rates per 1,000 populations. Public and private pawn forfeiture amounts are transformed using the inverse hyperbolic sine. Monetary values are real values adjusted by the wholesale price index, with December 1929 as the base period. All specifications include the full set of control variables, ward fixed effects, year fixed effects, and ward-specific linear trends. Control variables are the logarithm of population, male share of the population, business-taxpayer rate, the income-taxpayer rate, social worker rate, and police officers per 1,000 populations. Standard errors clustered at the ward level are reported in parentheses. The reported p-values test the null hypothesis that the pawnshop coefficients are equal across theft, fraud and extortion, and embezzlement.

Table 8: Private Pawn Loan Counts by Amount Group and Property-Related Crime Types

| | Theft | Fraud/Extortion | Embezzlement |
|---|---|---|---|
| | (1) | (2) | (3) |
| Public loan count | -0.208 | 0.148* | -0.023 |
| | (0.170) | (0.073) | (0.031) |
| Private loan count | -3.382 | -1.049* | -0.725** |
| × Below public maximum | (3.497) | (0.537) | (0.265) |
| Private loan count | -3.488 | -1.051* | -0.725** |
| × Above public maximum | (3.499) | (0.538) | (0.266) |
| All control variables | Yes | Yes | Yes |
| Ward FE | Yes | Yes | Yes |
| Year FE | Yes | Yes | Yes |
| Time trend | Yes | Yes | Yes |
| Observations | 135 | 135 | 135 |

***, **, and * represent statistical significance at the 1%, 5%, and 10% levels, respectively.
Notes: The dependent variables are crime rates per 1,000 populations. Public and private pawn loan counts are transformed using the inverse hyperbolic sine. The private loan count variables divide private pawnshop loan counts according to whether the average private pawn loan amount per transaction in a ward-year is at or below, or above, 8.5 yen, the maximum average loan amount observed for public pawnshops. All specifications include the full set of control variables, ward fixed effects, year fixed effects, and ward-specific linear trends. Control variables are the logarithm of population, male share of the population, business-taxpayer rate, income-taxpayer rate, social worker rate, and police officers per 1,000 populations. Standard errors clustered at the ward level are reported in parentheses.

Table 9: Implied changes in crime and monetized damage, 1928–1935

| Crime | Coefficient | Implied change in crimes | % of reported offenses | Damage-per -case (yen) | Damage averted (1,000 yen) |
|---|---|---|---|---|---|
| Panel A. Public pawnshops | | | | | |
| Theft | -0.276* | -5,469.60 | -1.11 | 36.08 | 185.44 |
| Fraud/Ext | 0.135*** | 2,675.35 | 4.09 | 316.62 | -646.75 |
| Embezzlement | -0.009 | -178.36 | -0.56 | 391.00 | 67.50 |
| Theft + fraud/ext | | -2,794.25 | -0.50 | | -461.30 |
| Total | | -2,972.61 | -0.50 | | -393.81 |
| Panel B. Private pawnshops | | | | | |
| Theft | -0.487 | -1,416.13 | -0.29 | 36.08 | 49.43 |
| Fraud/Ext | -1.443** | -4,196.06 | -6.42 | 316.62 | 1,140.99 |
| Embezzlement | -0.967*** | -2,811.91 | -8.76 | 391.00 | 1,062.40 |
| Fraud/ext + embezz | | -7,007.97 | -7.19 | | 2,203.39 |
| Total | | -8,424.10 | -1.42 | | 2,252.82 |

Notes: The sample covers the 15 wards of Tokyo from 1928 to 1935, using 1927 as the baseline year. The implied change in crimes is calculated from the change in IHS pawnshop lending relative to 1927, the corresponding coefficient, and ward population. Damage-per-case is calculated from year-specific Tokyo crime statistics for the fifteen wards and is deflated to December 1929 prices. Damage averted is computed at the ward-year level and summed across ward-years; positive values indicate implied damage avoided, while negative values indicate implied additional damage. Theft + fraud/ext and fraud/ext + embezz include coefficients statistically significant at the 10 percent level or better. Total uses all three point estimates.

Sources: Author’s calculations based on Table 4 and the data sources listed in Table 2.

# Appendices

## Appendix A: Data sources and variable definitions

All control variables, together with the population used as the denominator for the crime and other rates, are taken from the Statistical Yearbooks of Tokyo City (Tokyo City Office 1929–1937), using the figures recorded therein. The income-taxpayer, business-taxpayer, and social-worker rates count payers of the income tax, business tax, and the number of district welfare commissioners (homen-iin), respectively.

The percentage of protection-requiring persons used in Appendix B.1 is the number of residents the City classified as protection-requiring, divided by ward population, taken from the social surveys of 1929, 1931, 1932, 1933, and 1935 (Tokyo City Office 1930, 1932, 1934; Tokyo City Social Welfare Bureau 1933, 1936); the mean for each ward over these survey years is used.

The per-case damage figures used in the Discussion are, for each property-related-crime category, the reported value of property lost (currency plus the value of goods and securities) divided by the number of cases (Tokyo City Office 1929–1937).

## Appendix B: Additional analyses and robustness checks

Unless otherwise noted, each table re-estimates the baseline specification of Table 4: the inverse hyperbolic sine of public and private pawnshop loan amounts, ward fixed effects, year fixed effects, ward-specific linear trends, and the full set of control variables, with standard errors clustered at the ward level. The estimation sample is the balanced panel of 15 wards over 1927–1935 (135 ward-year observations); specifications that use a one-period lag or exclude 1927 have 120 observations.

### Appendix B.1: Spatial spillovers, cross-ward heterogeneity, and identification

These checks ask whether the baseline associations survive cross-ward spillovers, vary with ward economic conditions, and depend on how the public coefficient is identified.

Table B.1. Pawnshop loans, neighboring wards' loans, and crime

| | Theft<br>(1) | Fraud/Extortion<br>(2) | Embezzlement<br>(3) |
|---|---|---|---|
| Public pawn loan amount | -0.382** | 0.127*** | -0.025 |
| | (0.146) | (0.038) | (0.017) |
| Private pawn loan amount | 0.904 | -1.414** | -0.837*** |
| | (4.091) | (0.652) | (0.276) |
| Neighbor avg. public loan (std.) | -1.579 | -0.216 | -0.325 |
| | (1.287) | (0.349) | (0.236) |
| Neighbor avg. private loan (std.) | 7.149*** | -0.240 | 0.288 |
| | (2.327) | (0.514) | (0.245) |
| All control variables | Yes | Yes | Yes |
| Ward FE | Yes | Yes | Yes |
| Year FE | Yes | Yes | Yes |
| Time trend | Yes | Yes | Yes |
| Observations | 135 | 135 | 135 |

Notes: The dependent variables are crime rates per 1,000 population. Public and private pawn loan amounts are transformed using the inverse hyperbolic sine. Each specification adds standardized spatial lags, the row-normalized averages of contiguous wards' public and private loan amounts, constructed from a border-contiguity ward adjacency matrix and standardized to zero mean and unit variance. Monetary values are real, adjusted by the wholesale price index (December 1929 base). All specifications include ward fixed effects, year fixed effects, ward-specific linear trends, and the full set of control variables (log population, male share, business-taxpayer rate, income-taxpayer rate, social-worker rate, and police officers per 1,000 population). Standard errors clustered at the ward level are reported in parentheses. ***, **, and * denote significance at the 1%, 5%, and 10% levels, respectively.

Pawnshop activity and crime in one ward may be related to conditions in neighboring wards, because credit and offenders can cross ward boundaries and adjacent wards share local economic shocks. To assess whether such spillovers contaminate the baseline estimates, the baseline specification is augmented with the row-normalized average of contiguous wards' public and private loan amounts, constructed from a border-contiguity adjacency matrix and standardized to zero mean and unit variance (standardized spatial lags). The own-ward coefficients are essentially unchanged from the baseline, which suggests that the within-ward associations do not simply reflect omitted conditions in neighboring wards. The spatial lags themselves are insignificant in five of the six cases; the only exception is a positive association between neighbors' private lending and own-ward theft. This suggests that broader local credit conditions and theft move together across ward boundaries, being consistent with the stolen goods market channel based on widespread private pawnshops. Overall, these results provide little evidence that cross-ward spillover confounds the baseline findings.

Table B.2. Heterogeneity in the pawnshop-crime relationship by ward economic characteristics

| | Theft<br>(1) | Fraud/Extortion<br>(2) | Embezzlement<br>(3) |
|---|---|---|---|
| Panel A. Income taxpayers | | | |
| Public pawn loan amount | 13.142*** | 0.179 | 0.474 |
| | (4.189) | (1.257) | (0.599) |
| Public loan × income taxpayers | -3.417*** | -0.013 | -0.122 |
| | (1.058) | (0.319) | (0.149) |
| Private pawn loan amount | -8.577 | -4.457* | -0.035 |
| | (7.891) | (2.320) | (0.729) |
| Private loan × income taxpayers | 2.204 | 0.852 | -0.267 |
| | (2.248) | (0.632) | (0.234) |
| Panel B. Protection-requiring persons | | | |
| Public pawn loan amount | -1.465 | -0.595** | -0.101 |
| | (1.710) | (0.275) | (0.171) |
| Public loan<br>× protection-requiring persons | 0.192 | 0.117** | 0.015 |
| | (0.281) | (0.046) | (0.029) |
| Private pawn loan amount | -4.621 | 0.258 | -1.146 |
| | (8.198) | (1.526) | (0.667) |
| Private loan<br>× protection-requiring persons | 0.610 | -0.295 | 0.025 |
| | (0.860) | (0.198) | (0.098) |
| All control variables | Yes | Yes | Yes |
| Ward FE | Yes | Yes | Yes |
| Year FE | Yes | Yes | Yes |
| Time trend | Yes | Yes | Yes |
| Observations | 135 | 135 | 135 |

Notes: The dependent variables are crime rates per 1,000 population. Public and private pawn loan amounts are transformed using the inverse hyperbolic sine. Each panel adds interactions of the loan variables with a time-invariant ward characteristic, the ward sample-period mean income-taxpayer rate (Panel A) and the ward sample-period mean percentage of protection-requiring persons (Panel B), each measured as a percentage of ward population, whose main effects are absorbed by the ward fixed effects. Because the moderator enters in level form, the loan coefficients give the association at a moderator value of zero. Monetary values are real, adjusted by the wholesale price index (December 1929 base). All specifications include ward fixed effects, year fixed effects, ward-specific linear trends, and the full set of control variables (log population, male share, business-taxpayer rate, income-taxpayer rate, social-worker rate, and police officers per 1,000 population). Standard errors clustered at the ward level are reported in parentheses. ***, **, and * denote significance at the 1%, 5%, and 10% levels, respectively.

To examine whether the baseline associations vary with ward economic conditions, the loan variables are interacted with two time-invariant ward characteristics: each ward's sample-period mean income-taxpayer rate (Panel A) and its sample-period mean percentage of protection-requiring persons (Panel B), each of which is measured as a percentage of ward population. Because these characteristics are time-invariant, their main effects are absorbed by the ward fixed effects, and only the interactions are identified. Since the moderator enters in level form, the reported loan coefficient measures the association at a moderator value of zero rather than at its sample mean, which is why, for example, the public coefficient for theft in Panel A is positive even though the baseline (sample-average) association is negative.

The interactions are statistically significant only for the public coefficients, and in directions consistent with the mechanisms emphasized in the main text. In Panel A, the interaction between public lending and the income-taxpayer rate for theft is negative and significant, thereby indicating that the theft-reducing association of public lending is more pronounced in higher-income wards, where public lending is likely to relax liquidity constraints of relatively limited low-income residents, and where the expected return on theft is high due to the large number of potential high-income targets. In Panel B, the interaction between public lending and the percentage of protection-requiring persons for fraud/extortion is positive and significant: the positive association between public lending and fraud/extortion strengthens precisely where economically vulnerable residents are more concentrated. Furthermore, this pattern is consistent with the victimization channel, under which extending credit to vulnerable households also makes them more attractive targets.

Table B.3. Pawnshop loans and crime, by public-pawnshop presence

| | Theft<br>(1) | Fraud/Extortion<br>(2) | Embezzlement<br>(3) |
|---|---|---|---|
| Panel A. Setup wards (ever had a public pawnshop) | | | |
| Public pawn loan amount | -0.426*** | 0.083** | 0.006 |
| | (0.112) | (0.029) | (0.027) |
| Private pawn loan amount | -4.403 | -1.767** | -0.326 |
| | (4.945) | (0.561) | (0.376) |
| Observations | 63 | 63 | 63 |
| Panel B. Non-setup wards | | | |
| Public pawn loan amount | — | — | — |
| Private pawn loan amount | 4.335 | -1.681 | -1.633** |
| | (6.092) | (2.410) | (0.632) |
| All control variables | Yes | Yes | Yes |
| Ward FE | Yes | Yes | Yes |
| Year FE | Yes | Yes | Yes |
| Time trend | Yes | Yes | Yes |
| Observations | 72 | 72 | 72 |

Notes: The dependent variables are crime rates per 1,000 population. Public and private pawn loan amounts are transformed using the inverse hyperbolic sine. Panel A uses setup wards (those that ever hosted a public pawnshop); Panel B uses non-setup wards, where public lending is identically zero; therefore, the public coefficient is not estimable and is reported as a dash. Monetary values are real, adjusted by the wholesale price index (December 1929 base). All specifications include ward fixed effects, year fixed effects, ward-specific linear trends, and the full set of control variables (log population, male share, business-taxpayer rate, income-taxpayer rate, social-worker rate, and police officers per 1,000 population). Standard errors clustered at the ward level are reported in parentheses. ***, **, and * denote significance at the 1%, 5%, and 10% levels, respectively.

Because public lending is zero in wards that never hosted a public pawnshop, the public coefficient in the baseline is identified only from those that did. To make this explicit, the sample is split into setup wards, which ever hosted a public pawnshop, and non-setup wards, which did not. In the setup wards, the estimates reproduce the baseline public pattern as well as the negative association between private lending and fraud/extortion, which indicates that such results do not arise from pooling structurally different wards. In the non-setup wards, public lending is identically zero, so the public coefficient cannot be estimated and is reported as a dash in Panel B; the private coefficients in these wards are estimated from a smaller subsample and are accordingly less precise.

Appendix B.2: Dynamics and reverse causality

These checks examine the timing of the credit-crime relationships and the possibility that crime drives lending rather than the reverse.

Table B.4. Lagged pawnshop loans and crime

| | One-period lag<br>(1) | Contemp. + lag<br>(2) |
|---|---|---|
| Panel A. Theft | | |
| Public pawn loan amount (*t*) | | -0.215** |
| | | (0.076) |
| Public pawn loan amount (*t*-1) | -0.280 | -0.185 |
| | (0.160) | (0.174) |
| Private pawn loan amount (*t*) | | -1.041 |
| | | (4.003) |
| Private pawn loan amount (*t*-1) | -2.574 | -2.229 |
| | (5.483) | (5.367) |
| Panel B. Fraud/Extortion | | |
| Public pawn loan amount (*t*) | | 0.117*** |
| | | (0.030) |
| Public pawn loan amount (*t*-1) | 0.071 | 0.030 |
| | (0.051) | (0.043) |
| Private pawn loan amount (*t*) | | -1.118 |
| | | (0.820) |
| Private pawn loan amount (*t*-1) | -0.870 | -0.472 |
| | (0.881) | (1.065) |
| Panel C. Embezzlement | | |
| Public pawn loan amount (*t*) | | -0.029** |
| | | (0.011) |
| Public pawn loan amount (*t*-1) | 0.006 | 0.022 |
| | (0.018) | (0.017) |
| Private pawn loan amount (*t*) | | -0.648* |
| | | (0.319) |
| Private pawn loan amount (*t*-1) | -0.280 | -0.057 |
| | (0.343) | (0.386) |
| All control variables | Yes | Yes |
| Ward FE | Yes | Yes |
| Year FE | Yes | Yes |
| Time trend | Yes | Yes |
| Observations | 120 | 120 |

Notes: The dependent variables are crime rates per 1,000 population. Public and private pawn loan amounts are transformed using the inverse hyperbolic sine. Column 1 uses the one-period lag of the loan variables; Column 2 includes contemporaneous loans and their one-period lag. Both columns have 120 observations (one year is lost to the lag). Monetary values are real, adjusted by the wholesale price index (December 1929 base). All specifications include ward fixed effects, year fixed effects, ward-specific linear trends, and the full set of control variables (log

population, male share, business-taxpayer rate, income-taxpayer rate, social-worker rate, and police officers per 1,000 population). Standard errors clustered at the ward level are reported in parentheses. ***, **, and * denote significance at the 1%, 5%, and 10% levels, respectively.

To examine the timing of the relationships and probe for delayed effects, each crime type is regressed on the one-period lag of the loan amounts and, in a second specification, on contemporaneous loans together with their one-period lag. These relationships are driven by the contemporaneous coefficients, which largely retain the baseline signs, whereas the lagged terms are individually insignificant. This pattern suggests that the relationships operate through contemporaneous, within-ward variation rather than through delayed adjustment, and supports reading the baseline coefficients as reflecting same-year co-movement.

Table B.5. Granger-type test: lagged crime and pawnshop loans

| | Theft<br>(1) | Fraud/Extortion<br>(2) | Embezzlement<br>(3) |
|---|---|---|---|
| Panel A. Dependent variable: IHS public loan amount | | | |
| Lagged own crime rate (*t*-1) | -0.006 | 0.149 | 0.609 |
| | (0.021) | (0.155) | (0.672) |
| Panel B. Dependent variable: IHS private loan amount | | | |
| Lagged own crime rate (*t*-1) | 0.001 | -0.008 | -0.018 |
| | (0.004) | (0.010) | (0.023) |
| All control variables | Yes | Yes | Yes |
| Ward FE | Yes | Yes | Yes |
| Year FE | Yes | Yes | Yes |
| Time trend | Yes | Yes | Yes |
| Observations | 120 | 120 | 120 |

Notes: The dependent variable is the inverse hyperbolic sine of the public (Panel A) or private (Panel B) pawn loan amount, regressed on the one-period lag of the own-ward crime rate (per 1,000 population) for the indicated crime type. The sample has 120 observations (one year is lost to the lag). Monetary values are real, adjusted by the wholesale price index (December 1929 base). All specifications include ward fixed effects, year fixed effects, ward-specific linear trends, and the full set of control variables (log population, male share, business-taxpayer rate, income-taxpayer rate, social-worker rate, and police officers per 1,000 population). Standard errors clustered at the ward level are reported in parentheses. ***, **, and * denote significance at the 1%, 5%, and 10% levels, respectively.

As a more direct check on reverse causality, the direction of the regression is reversed and whether past crime predicts current lending examined. Specifically, each loan variable is regressed on the one-period lag of the corresponding crime rate, retaining the baseline controls, fixed effects, and ward-specific trends. None of the resulting coefficients is statistically significant, which indicates that lagged crime does not predict current public or private lending. This mitigates the concern that the baseline associations reflect lending responding to prior crime, even though contemporaneous reverse causality cannot be ruled out.

Appendix B.3: Alternative outcomes, measures, and samples

This section examines the robustness of the baseline results by considering an alternative crime measure, finer crime categories, sample period, and scaling of credit.

Table B.6. Pawnshop loans and arrest rates

| | Theft<br>(1) | Fraud/Extortion<br>(2) | Embezzlement<br>(3) |
|---|---|---|---|
| Public pawn loan amount | -0.045 | 0.133*** | 0.010 |
| | (0.061) | (0.032) | (0.021) |
| Private pawn loan amount | -2.619 | -0.983 | -0.770* |
| | (1.613) | (0.671) | (0.425) |
| All control variables | Yes | Yes | Yes |
| Ward FE | Yes | Yes | Yes |
| Year FE | Yes | Yes | Yes |
| Time trend | Yes | Yes | Yes |
| Observations | 135 | 135 | 135 |

Notes: The dependent variables are arrest rates per 1,000 population; arrest counts are from the Tokyo Metropolitan Police Department (1928–1936). Public and private pawn loan amounts are transformed using the inverse hyperbolic sine. Monetary values are real, adjusted by the wholesale price index (December 1929 base). All specifications include ward fixed effects, year fixed effects, ward-specific linear trends, and the full set of control variables (log population, male share, business-taxpayer rate, income-taxpayer rate, social-worker rate, and police officers per 1,000 population). Standard errors clustered at the ward level are reported in parentheses. ***, **, and * denote significance at the 1%, 5%, and 10% levels, respectively.

As an alternative measure of crime, reported-offense rates is replaced with arrest rates, which capture the subset of offenses that resulted in an arrest. The signs of the baseline associations are preserved. The associations between public lending and fraud/extortion and between private lending and embezzlement remain statistically significant, while those of public lending to theft and private lending to fraud/extortion attenuate to statistical insignificance. Some attenuation is expected, because arrest rates reflect not only the incidence of crime but also clearance and enforcement intensity, which add noise and may respond to different factors. Thus, the directions of persistence of the baseline associations support the robustness of the main findings. The data on arrests are obtained from the Tokyo Metropolitan Police Department (1928–1936).

Table B.7. Pawnshop loans and disaggregated property-related crime

| | Theft (excl. pickpocket) | Pickpocket | Fraud | Extortion |
|---|---|---|---|---|
| | (1) | (2) | (3) | (4) |
| Public pawn loan amount | -0.281*** | 0.009 | 0.083*** | 0.044*** |
| | (0.092) | (0.015) | (0.027) | (0.012) |
| Private pawn loan amount | -1.714 | -0.270 | -1.123* | -0.143 |
| | (4.344) | (0.580) | (0.585) | (0.294) |
| All control variables | Yes | Yes | Yes | Yes |
| Ward FE | Yes | Yes | Yes | Yes |
| Year FE | Yes | Yes | Yes | Yes |
| Time trend | Yes | Yes | Yes | Yes |
| Observations | 120 | 120 | 120 | 120 |

Notes: The dependent variables are disaggregated crime rates per 1,000 population: theft excluding pickpocketing, pickpocketing, fraud, and extortion. Because these categories are reported separately only from 1928, 1927 is dropped, leaving 120 observations. Public and private pawn loan amounts are transformed using the inverse hyperbolic sine. Monetary values are real, adjusted by the wholesale price index (December 1929 base). All specifications include ward fixed effects, year fixed effects, ward-specific linear trends, and the full set of control variables (log population, male share, business-taxpayer rate, income-taxpayer rate, social-worker rate, and police officers per 1,000 population). Standard errors clustered at the ward level are reported in parentheses. ***, **, and * denote significance at the 1%, 5%, and 10% levels, respectively.

To check that the crime-type heterogeneity is not an artifact of how the categories are aggregated, these are disaggregated further by splitting theft into ordinary theft and pickpocketing and separating fraud and extortion into their two components. The negative association between public lending and theft is driven by ordinary theft rather than by pickpocketing, and the positive public association holds for both fraud and extortion taken individually. Therefore, the contrasting public associations across crime types reflect genuine differences between offense types rather than the particular grouping used in the baseline.

Table B.8. Pawnshop loans and crime, 1928–1935 subsample

| | Theft (1) | Fraud/Extortion (2) | Embezzlement (3) |
|---|---|---|---|
| Public pawn loan amount | -0.272** | 0.127*** | -0.022* |
| | (0.092) | (0.036) | (0.012) |
| Private pawn loan amount | -1.984 | -1.265* | -0.652** |
| | (4.352) | (0.630) | (0.273) |
| All control variables | Yes | Yes | Yes |
| Ward FE | Yes | Yes | Yes |
| Year FE | Yes | Yes | Yes |
| Time trend | Yes | Yes | Yes |
| Observations | 120 | 120 | 120 |

Notes: The dependent variables are crime rates per 1,000 population. The sample drops 1927, the year whose crime categories were re-aggregated, leaving 120 observations for 1928–1935. Public and private pawn loan amounts are transformed using the inverse hyperbolic sine. Monetary values are real, adjusted by the wholesale price index (December 1929 base). All specifications include ward fixed effects, year fixed effects, ward-specific linear trends, and the full set of control variables (log population, male share, business-taxpayer rate, income-taxpayer rate, social-worker rate, and police officers per 1,000 population). Standard errors clustered at the ward level are reported in parentheses. ***, **, and * denote significance at the 1%, 5%, and 10% levels, respectively.

Because the crime categories were reported on a different basis in 1927 and had to be re-aggregated for consistency, the baseline is estimated on the 1928–1935 subsample, which retains only the years reported on a uniform basis. Dropping 1927 leaves the baseline pattern intact and, if anything, slightly sharper. This suggests that the results are not driven by the 1927 classification change or re-aggregation procedure.

Table B.9. Per-capita pawnshop measures and crime

| | Theft<br>(1) | Fraud/Extortion<br>(2) | Embezzlement<br>(3) |
|---|---|---|---|
| Panel A. Loan count per capita | | | |
| Public loan count per capita | 3.734 | 1.954 | 0.869 |
| | (12.076) | (3.690) | (1.742) |
| Private loan count per capita | -2.673 | -0.755** | -0.392** |
| | (2.811) | (0.328) | (0.146) |
| Panel B. Loan amount per capita | | | |
| Public loan amount per capita | 2.594 | 0.593 | 0.276 |
| | (2.448) | (0.492) | (0.312) |
| Private loan amount per capita | -0.117 | -0.097 | -0.090*** |
| | (0.403) | (0.062) | (0.024) |
| All control variables | Yes | Yes | Yes |
| Ward FE | Yes | Yes | Yes |
| Year FE | Yes | Yes | Yes |
| Time trend | Yes | Yes | Yes |
| Observations | 135 | 135 | 135 |

Notes: The dependent variables are crime rates per 1,000 population. The pawnshop measures are entered per capita and untransformed (in levels): loan counts per capita (Panel A) and loan amounts per capita (Panel B), replacing the baseline inverse-hyperbolic-sine loan totals. Monetary values are real, adjusted by the wholesale price index (December 1929 base). All specifications include ward fixed effects, year fixed effects, ward-specific linear trends, and the full set of control variables (log population, male share, business-taxpayer rate, income-taxpayer rate, social-worker rate, and police officers per 1,000 population). Standard errors clustered at the ward level are reported in parentheses. ***, **, and * denote significance at the 1%, 5%, and 10% levels, respectively.

The pawnshop measures are rescaled by population, using loan counts and amounts per capita. The negative associations of private lending with fraud/extortion and with embezzlement remain intact under this rescaling, whereas the public coefficients become statistically insignificant. This contrast is consistent with the small and spatially concentrated scale of public lending, for which ward-level totals are more informative than per-capita intensity, and reinforces the choice to measure credit provision as ward-level totals rather than per capita.

Appendix B.4: Inference and specification sensitivity

These checks assess the sensitivity of the results to the method of statistical inference and choice of trend and loan transformation.

Table B.10. Inference robustness: alternative standard errors and wild bootstrap

| | Theft | Fraud/Extortion | Embezzlement |
|---|---|---|---|
| | (1) | (2) | (3) |
| Public pawn loan amount | -0.276* | 0.135*** | -0.009 |
| *Cluster-robust SE* | (0.139) | (0.034) | (0.019) |
| *Driscoll–Kraay SE* | (0.081) | (0.038) | (0.033) |
| *Two-way SE* | (0.126) | (0.039) | (0.028) |
| *Wild bootstrap p-value* | [0.888] | [0.355] | [0.736] |
| Private pawn loan amount | -0.487 | -1.443** | -0.967*** |
| *Cluster-robust SE* | (4.837) | (0.613) | (0.241) |
| *Driscoll–Kraay SE* | (2.785) | (0.616) | (0.237) |
| *Two-way SE* | (4.667) | (0.593) | (0.308) |
| *Wild bootstrap p-value* | [0.920] | [0.042] | [0.011] |
| All control variables | Yes | Yes | Yes |
| Ward FE | Yes | Yes | Yes |
| Year FE | Yes | Yes | Yes |
| Time trend | Yes | Yes | Yes |
| Observations | 135 | 135 | 135 |

Notes: The dependent variables are crime rates per 1,000 population. Public and private pawn loan amounts are transformed using the inverse hyperbolic sine. Four inference methods are reported beneath each coefficient (labeled at the left of each block): cluster-robust, Driscoll–Kraay, and two-way (ward and year) standard errors in parentheses, followed by the wild-cluster-bootstrap p-value in brackets, computed using 999 replications. Monetary values are real, adjusted by the wholesale price index (December 1929 base). All specifications include ward fixed effects, year fixed effects, ward-specific linear trends, and the full set of control variables (log population, male share, business-taxpayer rate, income-taxpayer rate, social-worker rate, and police officers per 1,000 population). ***, **, and * denote significance at the 1%, 5%, and 10% levels, respectively.

As the panel data contain only 15 wards, conventional cluster-robust inference may be somewhat unreliable. Therefore, four alternative inference methods are reported beneath each coefficient, labeled at each row: cluster-robust (baseline), Driscoll–Kraay, and two-way (ward and year) clustered standard errors in parentheses, followed by the wild-cluster-bootstrap p-value in brackets. The negative associations of private lending with fraud/extortion and with embezzlement remain statistically significant across all four methods. By contrast, the public associations lose significance under the wild-cluster bootstrap. This discrepancy may reflect the sensitivity of inference for public pawnshops to the limited number of wards contributing identifying variation, a structure that resembles the few-treated-clusters problem in which wild-cluster-bootstrap procedures can perform unreliably (MacKinnon and Webb 2018). The evidence for statistical significance is therefore more tentative for public pawnshops than for private ones.

Table B.11. Pawnshop loans and crime under alternative fixed-effects and trend specifications

| | (1) | (2) | (3) | (4) |
|---|---|---|---|---|
| Panel A. Theft | | | | |
| Public pawn loan amount | -0.447*** | -0.276* | -0.175 | -0.195 |
| | (0.146) | (0.139) | (0.105) | (0.128) |
| Private pawn loan amount | 3.362 | -0.487 | -0.939 | 1.732 |
| | (3.601) | (4.837) | (3.776) | (4.259) |
| Panel B. Fraud/Extortion | | | | |
| Public pawn loan amount | 0.007 | 0.135*** | 0.070* | 0.101*** |
| | (0.055) | (0.034) | (0.038) | (0.026) |
| Private pawn loan amount | -1.417 | -1.443** | -1.301 | -1.260* |
| | (0.851) | (0.613) | (0.847) | (0.704) |
| Panel C. Embezzlement | | | | |
| Public pawn loan amount | 0.005 | -0.009 | -0.053*** | -0.013 |
| | (0.022) | (0.019) | (0.018) | (0.016) |
| Private pawn loan amount | -0.676** | -0.967*** | -0.724 | -1.242*** |
| | (0.243) | (0.241) | (0.424) | (0.406) |
| All control variables | Yes | Yes | Yes | Yes |
| Ward FE | Yes | Yes | Yes | Yes |
| Year FE | Yes | Yes | Yes | Yes |
| Ward linear trend | No | Yes | Yes | Yes |
| Ward quadratic trend | No | No | Yes | No |
| 1927 outcome × year | No | No | No | Yes |
| Observations | 135 | 135 | 135 | 135 |

Notes: The dependent variables are crime rates per 1,000 population. Public and private pawn loan amounts are transformed using the inverse hyperbolic sine. The columns vary the trend structure: (1) ward and year fixed effects only; (2) adds a ward-specific linear trend; (3) adds a ward-specific quadratic trend (which also includes the linear term); and (4) adds the 1927 outcome interacted with a year trend. The Yes/No rows indicate the components included in each column. Monetary values are real, adjusted by the wholesale price index (December 1929 base). All specifications include ward fixed effects, year fixed effects, and the full set of control variables (log population, male share, business-taxpayer rate, income-taxpayer rate, social-worker rate, and police officers per 1,000 population). Standard errors clustered at the ward level are reported in parentheses. ***, **, and * denote significance at the 1%, 5%, and 10% levels, respectively.

The baseline includes ward-specific linear trends to absorb divergent medium-run trajectories across wards. To assess sensitivity to this choice, this study varies the trend structure across columns: ward and year fixed effects only; adding a ward-specific linear trend (baseline); adding a ward-specific quadratic trend, which also includes the linear term; and adding the 1927 outcome interacted with a year trend. The key signs are stable across these specifications, even though statistical significances are not stable partially because of small sample size. These results suggest that the main findings are not driven by the specific estimation model.

Table B.12. Pawnshop loans and crime under alternative loan transformations

| | IHS<br>(1) | ln(*x*+1)<br>(2) | ln(*x*), zeros dropped<br>(3) |
|---|---|---|---|
| Panel A. Theft | | | |
| Public pawn loan amount | -0.276* | -0.289* | -3.728** |
| | (0.139) | (0.153) | (1.212) |
| Private pawn loan amount | -0.487 | -0.490 | -4.763 |
| | (4.837) | (4.841) | (4.451) |
| Panel B. Fraud/Extortion | | | |
| Public pawn loan amount | 0.135*** | 0.143*** | 0.250 |
| | (0.034) | (0.037) | (0.246) |
| Private pawn loan amount | -1.443** | -1.446** | -1.689*** |
| | (0.613) | (0.613) | (0.416) |
| Panel C. Embezzlement | | | |
| Public pawn loan amount | -0.009 | -0.009 | 0.220 |
| | (0.019) | (0.021) | (0.162) |
| Private pawn loan amount | -0.967*** | -0.967*** | -0.413 |
| | (0.241) | (0.241) | (0.283) |
| All control variables | Yes | Yes | Yes |
| Ward FE | Yes | Yes | Yes |
| Year FE | Yes | Yes | Yes |
| Time trend | Yes | Yes | Yes |
| Observations | 135 | 135 | 60 |

Notes: The dependent variables are crime rates per 1,000 population. The columns vary the transformation of the public and private loan amounts: (1) inverse hyperbolic sine (baseline); (2) ln(*x* + 1); and (3) ln(*x*) with zero-public-lending ward-years dropped (60 observations remaining). Monetary values are real, adjusted by the wholesale price index (December 1929 base). All specifications include ward fixed effects, year fixed effects, ward-specific linear trends, and the full set of control variables (log population, male share, business-taxpayer rate, income-taxpayer rate, social-worker rate, and police officers per 1,000 population). Standard errors clustered at the ward level are reported in parentheses. ***, **, and * denote significance at the 1%, 5%, and 10% levels, respectively.

In the baseline analyses, the loan variables are transformed using the inverse hyperbolic sine, which accommodates the zero values present in public lending. The baseline transformation is also compared with two alternatives: ln(*x* + 1), which similarly accommodates zeros, and ln(*x*) with zero-lending ward-years dropped. The inverse hyperbolic sine and ln(*x* + 1) yield virtually identical estimates. The ln(*x*) specification discards all ward-years with zero public lending and so is not directly comparable, but the negative association between private lending and fraud/extortion persists even in the restricted sample. The baseline transformation is therefore not driving the results.

Appendix References

Tokyo City Office. (1929–1937). *Tokyoshi tokei nenpyo, dai 25–33 kai [Statistical Yearbooks of Tokyo City, Volumes 25–33]*. Tokyo.

Tokyo City Office. (1930). *Tokyoshinai yohogosha ni kansuru chosa [Survey of protection-required people in Tokyo City]*. Tokyo.

Tokyo City Office. (1932). *Tokyoshi yohogosetai seikei chosa [Living survey of protection-required households in Tokyo City]*. Tokyo.

Tokyo City Office. (1934). *Tokyoshi yohogosetai chosa [Survey of protection-required households in Tokyo City]*. Tokyo.

Tokyo City Social Welfare Bureau. (1933). *Tokyoshi yohogosha chosa [Survey of protection-required people in Tokyo City]*. Tokyo.

Tokyo City Social Welfare Bureau. (1936). *Tokyoshi shakaikyoku nenpo, dai 16 kai [Annual reports of the Tokyo City Social Welfare Bureau, volume 16]*. Tokyo.

Tokyo Metropolitan Police Department. (1928–1939, 1941–1942). *Keishicho tokeisho, dai 37–50 kai [Statistical Reports of the Tokyo Metropolitan Police Department, volumes 37–50]*. Tokyo.

MacKinnon, James G., and Matthew D. Webb. (2018). "The Wild Bootstrap for Few (Treated) Clusters." *The Econometrics Journal* 21(2): 114–35.